\documentclass[aoas,MSNbibl,nameyear,seceqn,dvips]{arximspdf}
\usepackage{algorithm}
\usepackage{algpseudocode}
\usepackage{dcolumn}
\usepackage{graphicx}

\doi{10.1214/13-AOAS696} 
\volume{8}
\issue{1}
\pubyear{2014}
\firstpage{448}
\lastpage{480}

\makeatletter

\newcolumntype{d}[1]{D{.}{.}{#1}}

\newcommand{\rright}{\right}
\newcommand{\lleft}{\left}
\newcommand{\rrVert}{\Vert}
\newcommand{\llVert}{\Vert}
\makeatother

\begin{document}
\begin{frontmatter}

\title{Reconstructing evolving signalling networks by~hidden Markov
nested effects models\thanksref{T1}}
\runtitle{Hidden Markov nested effects models}

\begin{aug}
\author[A]{\fnms{Xin} \snm{Wang}\ead[label=e1]{xinwang2hms@gmail.com}},
\author[A]{\fnms{Ke} \snm{Yuan}\ead[label=e3]{ke.yuan@cruk.cam.ac.uk}},
\author[A]{\fnms{Christoph} \snm{Hellmayr}\ead[label=e2]{ch.hellmayr@gmail.com}},
\author[A]{\fnms{Wei} \snm{Liu}\ead[label=e5]{wei.liu@cruk.cam.ac.uk}}\\
\and
\author[A]{\fnms{Florian} \snm{Markowetz}\corref{}\ead[label=e4]{florian.markowetz@cruk.cam.ac.uk}\ead[label=u1,url]{http://www.markowetzlab.org/}}
\runauthor{X. Wang et al.}
\affiliation{University of Cambridge} 
\address[A]{Cancer Research UK Cambridge Institute\\
University of Cambridge\\
Li Ka Shing Centre\\
Robinson Way\\
Cambridge, CB2 0RE\\
United Kingdom\\
\printead{e1}\\
\phantom{E-mail:\ }\printead*{e3}\\
\phantom{E-mail:\ }\printead*{e4}\\
\printead{u1}} 
\end{aug}
\thankstext{T1}{Supported in part by The University of Cambridge,
Cancer Research UK and Hutchison Whampoa Limited.}

\received{\smonth{3} \syear{2013}}
\revised{\smonth{9} \syear{2013}}

%
\begin{abstract}
Inferring time-varying networks is important to understand the
development and
evolution of interactions over time. However, the vast majority of currently
used models assume direct measurements of node states, which are often
difficult to obtain, especially in fields like cell biology, where
perturbation experiments often only provide indirect information of network
structure. Here we propose hidden Markov nested effects models
(HM-NEMs) to
model the evolving network by a Markov chain on a state space of
signalling networks, which are derived from nested effects models
(NEMs) of indirect perturbation data. To infer the hidden network evolution
and unknown parameter, a Gibbs sampler is developed, in which sampling network
structure is facilitated by a novel structural Metropolis--Hastings algorithm.
We demonstrate the potential of HM-NEMs by simulations on synthetic
time-series perturbation data. We also show the applicability of
HM-NEMs in two
real biological case studies, in one capturing dynamic
crosstalk during the progression of neutrophil polarisation, and in the other
inferring an evolving network underlying early differentiation of mouse
embryonic stem cells.
\end{abstract}

%

\end{frontmatter}


\setcounter{footnote}{1}
\section{Introduction}\label{intro}
Understanding the inner workings of a complex system in biology,
ecology and
many other fields often relies on interventions to the system as well as
subsequent observation and analysis of perturbation effects. For
example, in
biology, gene silencing such as RNA interference (RNAi) followed by phenotypic
screening has become a widely used approach to study functions of genes
and the
signalling interactions between them [\citet{boutros2008art}].
Recently, there is
an increasing interest in generating and analysing time-series
phenotypic screens
after perturbations of a signalling network in order to study dynamics
of a
biological system over time [\citet{ivanova2006dissecting}, \citet{neumann2010phenotypic},
\citet{ku2012network}].
Despite some success in time-series analysis as well as graphical modelling,
there is still a lack of methodology specifically targeting this type
of data to
reconstruct time-varying signalling networks.

\subsection{Related work}
Recently, a lot of effort has been put into extending graphical models
to infer
evolving networks. In 2006, a generalised exponential random graph
model (ERGM)
was proposed to model the temporal progression of social networks
[\citet{hanneke2006discrete}]. ERGM was further extended to a hidden
temporal ERGM
(htERGM) by considering the networks as latent variables and using a local
kernel-weighting technique to learn smoothly evolving graphical
Gaussian models
[\citet{guo2007recovering}]. It was also proposed to model discrete time-varying
graphs as evolving Markov random fields and perform graphical
regression under
smoothness assumption with $l1$ shrinkage [\citet{ahmed2009recovering}].
Another major group of models in this field are dynamic Bayesian networks
(DBNs). The original DBN was developed under the assumption of a~homogeneous
Markov chain [\citet{murphy2002dynamic}]. Nonstationary continuous DBNs were
developed recently by assuming a fixed network with varying interaction
parameters [\citet{grzegorczyk2009non}]. Network structures are allowed
to change
in nonstationary DBNs for discrete data [\citet{robinson2009non}].
Nonhomogeneous DBNs were proposed to consider evolving networks as multiple
changepoint regression models and use reversible jump MCMC methods to do
inference [\citet{lebre2007stochastic}]. Further improvements of nonhomogeneous
DBNs were made by introducing information sharing among time series segments
[\citet{husmeierinter}]. Moreover, a time-varying DBN \mbox{(TV-DBN)} model
was proposed
to infer directed time-varying network structure by a kernel-weighting
$l1$-regularised auto-regressive approach [\citet{song2009time}].

These models have their own advantages and were all demonstrated to be effective
in specific applications. Their common limitation, however, is that they
reconstruct time-varying networks from direct observations of nodes,
which in
cell biology are often difficult to obtain. For example, measuring activities
of signalling proteins in a cellular pathway is very difficult, as they are
mostly mediated by post-translational modifications, which are usually not
visible in gene expression data, the most commonly used data for
network inference
[\citet{markowetz2010understand}].

\citet{markowetz2005non} addressed this problem by
introducing nested effect models (NEMs) to reconstruct signalling
networks from \mbox{observations} of downstream genes whose expression levels are affected by
perturbations of signalling proteins.
The name nested effect models stems from the fact that NEMs infer
directed relations
between signalling proteins by \emph{subset relations} between their
perturbation
effects [\citet{markowetz2007nested}].
Since their introduction [\citet{markowetz2005non}, \citet{markowetz2007nested}], static NEMs have
been extended in different directions and have been applied in several
case studies [\citeauthor{Froehlich2007} (\citeyear{Froehlich2007,Froehlich2008c}), \citet{Tresch2008},
\citet{vaske2009factor}, \citet{anchang2009modeling}, \citet{House2010},
\citet{frohlich2011fast}, \citet{niederberger2012mc}, \citet{SadehMS13},
\citet{failmezger2013learning}].

NEMs have been extended to model dynamics within signalling
pathways under the assumption that the observed dynamic perturbation effects
over time are due to time delay of signal transduction [\citet{anchang2009modeling}].
Another extension of NEMs infers time-varying networks by unrolling the network
structure over time [\citet{frohlich2011fast}, \citet{failmezger2013learning}].

\subsection{Contributions of this article}
We propose the hidden Markov nested effects models (HM-NEM) to
infer time-varying signalling networks. The evolving network is
modelled as a
discrete time first-order Markov process. The state space corresponds to
all possible network topologies. The transition probabilities of the
Markov process
are defined by a geometric distribution, which exploits the topological distance
(defined as the distance between their adjacency matrices) between networks.
Similar to the ``smoothness'' assumption in TV-DBNs, we assume that the more
distant two networks are, the less likely the transition is.
Importantly, such
``smoothness'' is controlled by a parameter which can be estimated from
the data.
For the observation model, NEMs provide the formulation of emission
probabilities that link the hidden network topologies to the observable
perturbation effects.

From a Bayesian perspective, the inference target is the joint posterior
distribution of the time-varying networks and the unknown smoothness parameter,
given the observed effects. To approach the target distribution, we
propose a
Gibbs sampler based on ``Metropolis-within-Gibbs''. The algorithm alternates
between sampling the state path and the parameter from the
corresponding full
conditionals.

In the next section, we briefly describe nested effects models and introduce
the marginal likelihood as well as inference methods.
The description and inference method of HM-NEM are presented in
Sections~\ref{hmnem} and \ref{mcmc}, respectively.
In Section~\ref{sim} we demonstrate simulation studies including convergence
diagnosis, sensitivity and coverage analysis on synthetic data for networks
with slow, moderate and rapid transitions.
Finally, we show two applications of HM-NEM to reconstruct the polarisation
network of neutrophils in Section~\ref{neutro} and the self-renewal signalling
network of embryonic stem cells in Section~\ref{esc}.

\section{Model}
\subsection{Background}
\label{nem}
Nested effects models (NEMs) are a statistical approach that is specifically
tailored to reconstruct features of pathways from perturbation effects in
downstream reporters [\citet{markowetz2007nested}]. In contrast
to other graphical models, which are all based on measures of pairwise
association (e.g., coexpression networks) and encode conditional independence
relations (e.g., Bayesian networks), NEMs describe \emph{subset relationships}
between observed downstream effects of perturbations.

\begin{figure}

\includegraphics{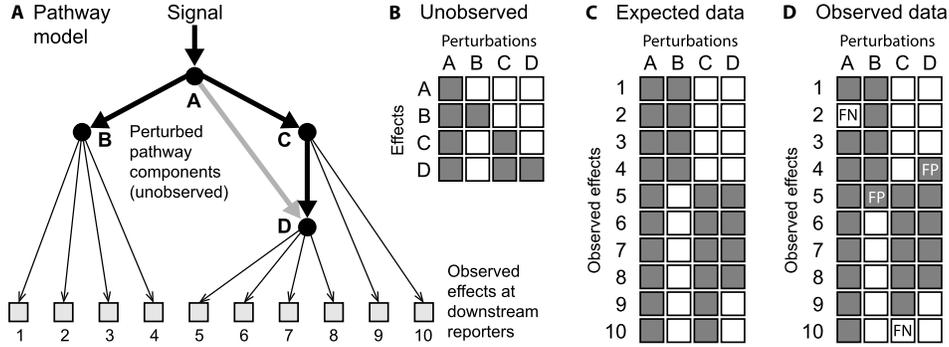}

\caption{A schematic figure of nested effects models.
\textup{(A)}~A toy example of NEMs. Perturbed pathway components (often called \emph{S-genes}, referring
to the {s}ignalling genes that are {s}ilenced by
perturbations) and effect reporters (called \emph{E-genes} if they are
transcriptional) are shown in black round circles and grey rectangles,
respectively.
The thick black arrows between pathway components denote their
signalling relationships,
while the thin black arrows connect pathway components to their direct
effect reporters.
The thick grey arrow between pathway components A and D is a transitive
edge indicating
an indirect effect of A on D (via C).
\textup{(B)}~States of pathway components upon different perturbations
are generally unobserved. A pathway component is in the state of 0
(white rectangle) if it is not perturbed, and in the state of 1 (black
rectangle) if it is perturbed directly or by a perturbation propagated
down from an upstream component.
\textup{(C)}~The expected states of effect reporters (black${}={}$effect;
white${}={}$no effect) after perturbing pathway components.
\textup{(D)}~In real biological applications, the noisy measurement of
effect reporter states may include false negatives (FNs) and false
positives (FPs).}\label{fig-nem}
\end{figure}

To describe NEMs more clearly, we show a toy signalling network
[Figure~\ref{fig-nem}(A)] consisting of a kinase (A) and three transcription
factors (B, C and D), which directly regulate reporter genes (1 to 10).
Phenotypic data generally do not include the states of proteins A to
D after perturbations [Figure~\ref{fig-nem}(B)]. This is because phenotypes
like gene expression or cell morphologies are downstream of the pathway
of interest and often do not contain much information on the activity states
of the proteins in the pathway.

NEMs assume that
perturbing upstream pathway components may impact a~global process,
while silencing
downstream genes only affect local subprocesses. This results in a subset
pattern in the observed data.
For example, for the pathway in Figure~\ref{fig-nem}(A) the
perturbation effects (e.g., expression changes in reporter genes)
of pathway components B, C and D are subsets of the effect of gene A:
perturbing~A has an
effect on all reporters (1--10), while perturbing~B only affects
reporters 1--4,
perturbing~C affects reporters 5--10, and perturbing~D affects reporters 5--8
[Figure~\ref{fig-nem}(C)].

NEMs leverage the observed data on effect reporters [Figure~\ref{fig-nem}(D)] and infer the most likely pathway structure that can
explain the subset patterns by comparing it to the data expected for a pathway
[Figure~\ref{fig-nem}(C)].
If the data are transcriptional phenotypes of RNAi experiments (like in
our second case study), the pathway components and effect reporters are
also called S-genes and E-genes [\citet{markowetz2005non}].

%

A pathway component $j$ is in the natural state of 0 ($S_{jk}=0$) if
it is not affected by perturbation $k$, and is in the state of 1
($S_{jk}=1$) if it is interrupted directly by biological perturbation or
indirectly by propagation of a perturbation from an upstream pathway component
[e.g., Figure~\ref{fig-nem}(B)]. In real biological applications, the
perturbation
can be either enhancing (e.g., by drug treatment in our application to
neutrophil
polarisation) or inhibiting (e.g., by RNAi in our application to mouse embryonic
stem cell differentiation). Since data sets generally contain either
one or the other of these perturbations, we do not distinguish between
enhancing and inhibiting perturbations in the description of our method.

A signalling network is
modelled by $G=(\mathcal{V}, \mathcal{E})$, in which $\mathcal{V}$
is the set of pathway components and $\mathcal{E}$ a set of all
interactions between
them. Let $D=[d_{ik}]_{m\times l}$ be the observed perturbation effects
[e.g., Figure~\ref{fig-nem}(D)], where $d_{ik}$ is the effect of
perturbation $k$
on effect reporter $i$.\vspace*{9pt}

\textit{Transitivity}. An important feature of NEMs is the
transitivity of
subset relations~[\citet{markowetz2007nested}]. For example, the perturbation
effect of gene D is a subset of the effect of gene C, which is a subset of
the effect of gene A [Figure~\ref{fig-nem}(C)]. Then the perturbation effect
of gene D must also be a subset of gene~A. Thus, the graph encoding such
transitive subset relations should be transitively closed: whenever
there is a path from one pathway component to another one (e.g.,
$A\rightarrow C\rightarrow D$),
a directed edge (e.g., $A\rightarrow D$) exists between these two
pathway components in the graph.\vspace*{9pt}

\textit{Marginal likelihood}.
Here, we make a convenience
assumption that each effect reporter is specific for only a single
pathway component.
Let $\Theta:= \{\theta_i\}_1^m$, where $\theta_i\in\{1, \ldots, n\}
$, be a set of
parameters indicating the positions of effect reporters. Reporter $i$ is
specific for component $j$ if $\theta_i=j$. Usually, reporter
positions are
unknown; thus, the likelihood of the signalling network $G$ given the
observation $D$ is computed by marginalisation over $\Theta$:
\begin{eqnarray}
\label{eq1} P(D\vert G) & = &\int P(D \vert G, \Theta)P(\Theta\vert G)
\,d\Theta
\nonumber\\[-8pt]\\[-8pt]
& = &\frac{1}{n^m} \prod_{i=1}^{m} \sum
_{j=1}^n \prod
_{k=1}^l P(d_{ik} \vert G,
\theta_i=j),\nonumber
\end{eqnarray}
where the first product is over all effect reporters under the
assumption that reporters are independent of each other, while the
second product is over
all replicates under the assumption that replicates are independent of each
other. During marginalisation, each effect reporter is ``attached'' to
all pathway components; we thus implicitly take multiple regulators
into account (but not complex interactions between them).

For a single effect reporter $i$ under perturbation $k$, the
probability to observe
$d_{ik}$ given $G$ and its position $\theta_i=j$ can be computed by
\begin{eqnarray}
\label{eq2} &&\hspace*{87.5pt} d_{ik}=1\hspace*{15.5pt} d_{ik}=0
\nonumber
\\[-8pt]
\\[-8pt]
&& P(d_{ik}\vert G,\theta_{i}=j)= \left\{\matrix{ \alpha &\quad 1-\alpha &\quad S_{jk}=0,
\cr
1-\beta&\quad \beta&\quad S_{jk}=1,}\right.\nonumber
\end{eqnarray}
where $\alpha$ and $\beta$ are global false positive and false
negative rate of
$D$ that can often be estimated from control experiments~[\citet{markowetz2005non}];
$S_{jk}$~is the state of pathway component $j$ upon perturbation $k$.

However, when the relationships between pathway components and effect
reporters are
already known, the likelihood can be simplified to
\begin{eqnarray}
\label{eq1b} P(D\vert G) & = &\prod_{i=1}^{m}
\prod_{k=1}^l P(d_{ik} \vert G).
\end{eqnarray}

\textit{Inference}.
The original NEM performs an exhaustive search over all transitively
closed graphs
to identify the optimal network by the maximum likelihood
estimation~[\citet{markowetz2005non}].
For large-scale networks consisting of many signalling genes,
heuristics such
as pairwise, triplets inference~[\citet{markowetz2007nested}] and
module networks [\citeauthor{Froehlich2007} (\citeyear{Froehlich2007,Froehlich2008c})]
have been developed.
The definition of effects has also been extended, for example, by modelling
differential expression as a mixture model of $p$-values~[\citet{Froehlich2007}]
or log-ratios~[\citet{Tresch2008}].
Moreover, \citet{niederberger2012mc} proposed an efficient
inference method by
combining MCMC sampling with an Expectation--Maximisation (EM) algorithm.

Maximum likelihood inference was adopted in the original NEM, while the model
itself can be naturally extended by incorporating priors of the signalling
network and/or the regulatory structure between signalling genes and reporter
genes.
\citet{Froehlich2007} extended NEMs to take into account prior knowledge
about network structure, and applied Bayesian regularisation using Akaike
information criterion (AIC). Maximum \textit{a posteriori} estimation
was also proposed to incorporate
prior network structure by \citet{Tresch2008} and \citet{Froehlich2008c}.

\subsection{Hidden Markov nested effects model}
\label{hmnem}
As mentioned in Section~\ref{intro}, NEMs establish a framework for
reconstruction of static signalling networks from perturbation effects.
However, NEMs and their extensions do not allow network structure to
change over
time in their present forms. Here, we extend NEM to the Hidden Markov nested
effects model (\textit{HM-NEM}) to model signalling networks with topological
changes over discrete time points.

\begin{figure}[t]
\includegraphics{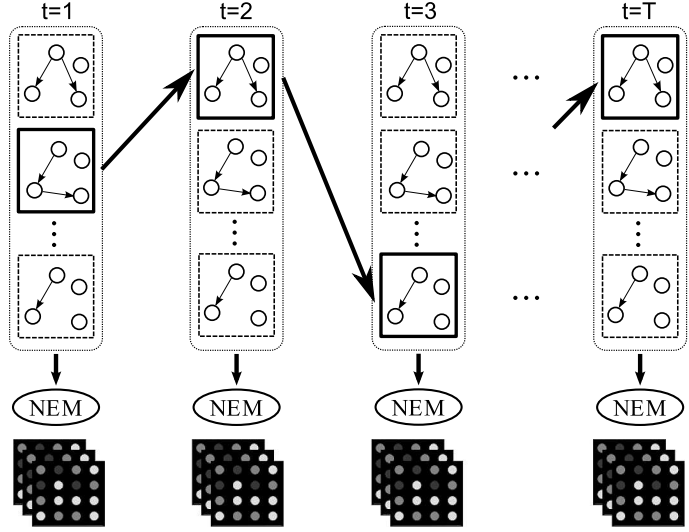}
\caption{A schematic figure describing the framework of HM-NEMs.
HM-NEMs are comprised of a Markov chain modelling an
evolving signalling network (the upper layer) and NEMs as the emission module
(the middle layer).
Images in the lower layer represent observed time-series perturbation effects.
All possible networks for a given number of pathway components are
included in
the state space denoted by dotted rounded rectangles.
At each time step, only one state is present and is highlighted by a
thick black
square.
The state transitions between adjacent time points are highlighted by thick
black arrows.}\label{fig-hmnem}
\end{figure}

The time-varying network is considered as a discrete stochastic process
$G_{1\dvtx T}=
\{G_t\}_1^T$. Let $G_t=(\mathcal{V}, \mathcal{E}_t)$ be the network at
time $t$ for $t\in\{1, \ldots, T\}$, in which $V$~is the set of
pathway components and
$\mathcal{E}_t$ is the edge set (throughout this paper edge means
directed edge) including all signalling interactions between pathway
components at
timestep $t$. Let $D_{1\dvtx T}= \{D_t\}_1^T$, where $D_t=[d_{ikt}]_{m\times
l}$ is a
matrix of observed effects for $m$ effect reporters across $l$ perturbations.
Under the first-order Markov assumption the probability of the
observation of
$G_t$ only depends on its previous network structure $G_{t-1}$ for
$t\in\{2,\ldots, T\}$ (the upper layer in Figure~\ref{fig-hmnem}).
%

A hidden Markov nested effects model (HM-NEM) is a hidden Markov model (HMM)
with the time-varying network as the transition module and nested
effects model
as the emission module. Similar to ordinary HMMs, we represent \mbox{a~HM-NEM} as
$\mathcal{H}_{\mathrm{nem}}=(\pi, A, B)$, in which $\pi$ is the initial
distribution over
states (network structures), and $A$ and $B$ denote the transition probabilities
and emission probabilities, respectively. Since the state space of the hidden
layer consists of \emph{all} possible network structures, no model selection
is needed to determine the number of states, which is fixed but can be very
large (e.g., \mbox{$>$1} billion states for a network consisting of only six pathway
components).\vspace*{9pt}

\textit{Initial distribution}.
The initial distribution is set to be a uniform distribution over all the
possible network structures.
This setting makes the initial distribution take no part in the inference
methods, reflecting the fact that prior information about the underlying
network is often not available in gene perturbation studies.\vspace*{9pt}

\textit{Transition probability}.
We assume here that a signalling network prefers to transit to a state
with a
similar structure. That is, let $P_{uv}$ be the probability to transit from
state $u$ to $v$, then the more distant the network structure $v$ is
from~$u$, the lower the transition probability $P_{uv}$ will be. This assumption
is sound
in biology, and many other graphical models for modelling time-varying networks
also make similar assumptions [e.g., TV-DBN in \citet{song2009time}
and TESLA in~\citet{ahmed2009recovering}].
Here we use the geometric distribution to derive the transition probabilities:
\begin{eqnarray}
\label{eq3} P_{uv} & =&P(G_{t+1}=v \vert G_t =
u)=\frac{1}{Z_u} (1-\lambda)^{s_{uv}} \lambda,
\end{eqnarray}
where $Z_u=\sum_u (1-\lambda)^{s_{uv}} \lambda$ is a normalising
constant; $\lambda\in(0,1)$ is a parameter controlling the ``smoothness'';
$s_{uv}$ is the distance between network $u$ and $v$ computed by
\begin{eqnarray}
\label{eq4} s_{uv} & = &\bigl\| A^u - A^v
\bigr\|_1:= \sum_{r}\sum _{c} \bigl| a^u_{rc} -a^v_{rc} \bigr|,
\end{eqnarray}
where $A^u$ and $A^v$ are binary adjacency matrices of networks $u$ and
$v$; precisely, $a_{rc} = 1$ denotes a directed edge from vertice $r$ to
$c$, and $0$ otherwise. Each $G_t$ corresponds to a unique adjacency matrix
$A_t$. In the following we show that this transition model
is a special case of the htERGM in \citet{guo2007recovering}.\vspace*{9pt}

\textit{Connection with the htERGM} [\textit{\citeauthor{guo2007recovering}} (\citeyear{guo2007recovering})].
The transition probability setting for the htERGM is
\begin{eqnarray}
P(G_{t+1}|G_{t}) &=& \frac{1}{Z(\bolds{\theta}, G_{t})} \exp\bigl\{ \bolds{
\theta}^\mathrm{T}\bolds{\Psi}(G_{t+1}, G_{t}) \bigr\},
\end{eqnarray}
where $ \bolds{\Psi}(G_t,G_{t-1})$ is a vector-valued function
corresponding to several
statistics of the two network topologies. The parameter $\bolds{\theta}$
contains the
weights of these features.

To see that our transition model is a special case of the above, we rewrite
equation~(\ref{eq3}) with the terminology of \citet{guo2007recovering}:
\begin{eqnarray*}
P(G_t|G_{t-1}) &=& \frac{1}{Z(\lambda, G_t)} \exp\biggl( \sum
_r \sum_c
\bigl|a_{rc}^{G_{t+1}} - a_{rc}^{G_{t}} \bigr| \ln(1 -
\lambda) + \ln\lambda\biggr)
\\
&=& \frac{1}{Z(\lambda, G_t)} \exp\lleft(\rule{0pt}{20pt}\matrix{ \bigl[\ln\lambda, \ln(1
-\lambda) \bigr]
\lleft[\matrix{ 1
\vspace*{4pt}\cr
\displaystyle \sum_r \sum
_c \bigl|a_{rc}^{G_{t+1}} - a_{rc}^{G_{t}}
\bigr| } \rright] }\rright)
\\
&=& \frac{1}{Z(\bolds{\theta}, G_t)} \exp\bigl(\bolds{\theta}^{\mathrm{T}}
\bolds{
\Phi}(G_{t+1}, G_t) \bigr),
\end{eqnarray*}
where
\begin{eqnarray*}
\bolds{\theta} &=& \lleft[\matrix{\ln\lambda
\cr
\ln(1 -\lambda) } \rright], \qquad\bolds{\Psi} = \lleft[\matrix{1
\vspace*{4pt}\cr
\displaystyle\sum_r \sum_c
\bigl|a_{rc}^{G_{t+1}} - a_{rc}^{G_{t}} \bigr|} \rright],
\\
Z(\bolds{\theta}, G_t) &=& \sum_{G_{t+1}} \exp
\bigl(\bolds{\theta}^\mathrm{T} \bolds{\Phi}(G_{t+1}, G_t)
\bigr).
\end{eqnarray*}

Note that the second dimension of the feature vector $\bolds{\Psi}$
directly tells
the amount of changes between the two networks. Its corresponding
weight parameter
$\ln(1-\lambda)$ can be used to generate/fit difference levels of network
dynamics. This is consistent with our design for $\lambda$ being the
probability of
an edge staying the same.\vspace*{9pt}

\textit{Emission probability}. The emission probability in a HM-NEM
is the
probability to observe perturbation effects $D_t$ at time $t$ given the current
network topology~$G_t$, which can be derived directly from the marginal
likelihood of the nested effects model in equation~(\ref{eq1}) when the
relationships of pathway components to effect reporters are unknown:
\begin{eqnarray}
\label{eq5} P(D_t\vert G_t)&=&\frac{1}{n^m}\prod
_{i=1}^{m}\sum_{j=1}^{n}
\prod_{k=1}^{l}P(d_{ikt}\vert
G_t, \theta_i=j).
\end{eqnarray}
Similarly, if the relationships between pathway components and effect
reporters are already
known, the emission probability can be derived from equation~(\ref{eq1b}):
\begin{eqnarray}
\label{eq5b} P(D_t\vert G_t)&=&\prod
_{i=1}^{m}\prod_{k=1}^{l}P(d_{ikt}
\vert G_t, \theta_i=j).
\end{eqnarray}
%

\section{Inference}\label{mcmc}
Having established the framework of HM-NEMs, our main interest is to
infer the
joint posterior distribution $P(G_{1\dvtx T},\lambda
\vert D_{1\dvtx T})$ of state sequence $G_{1\dvtx T}$
and the unknown parameter $\lambda$ in equation~(\ref{eq3}) given the
phenotype of
gene perturbations over time $D_{1\dvtx T}$. To approach this target
distribution, we use a Gibbs sampler which
draws samples from the two full conditionals: $P(G_{1\dvtx T}\vert\lambda,D_{1\dvtx T})$ and
$P(\lambda\vert G_{1\dvtx T}, D_{1\dvtx T})$.\vspace*{9pt}

\textit{Sampling $G_{1 \dvtx T}$}.
The full conditional of states is obtained by a single-site-update approach,
which samples one hidden state at a time. As such, letting $G_{-t}:=\{G_{t^\prime}; t^\prime\neq t\}$, the target distribution becomes the
conditional distribution of each state given all the other states, data and
parameter, which can be written as 
\begin{eqnarray*}
&& P(G_t=s \vert G_{-t}, D_{1\dvtx T}, \lambda)
\\
&&\qquad \propto\cases{ P(G_{t+1}\vert G_{t}=s, \lambda)P(D_{t}\vert G_t=s),\qquad \mbox{if }t=1,
\vspace*{3pt}\cr
P(G_t=s\vert G_{t-1}, \lambda)P(G_{t+1}\vert G_{t}=s, \lambda) P(D_{t}\vert G_t=s),
\vspace*{3pt}\cr
\hspace*{175pt}\mbox{if }t = 2,\ldots,T-1,
\vspace*{3pt}\cr
P(G_t=s\vert G_{t-1}, \lambda)P(D_{t}\vert G_t=s),\qquad\mbox{if }t=T.}
\end{eqnarray*}

Direct sampling from this distribution is infeasible. Hence, we
resort to the Metropolis-within-Gibbs approach [\citet{geyer2010introduction}]
which facilitates sampling by the Metropolis--Hastings (MH) algorithm.

To sample networks, we propose a structural MH. By contrast to the
method in
\citet{madigan1995bayesian}, this MH does not restrict the state
space to DAGs.
In detail, we use a uniform jumping distribution to propose new graphs: a~new
graph $s^\prime$ is generated by adding or deleting an edge selected randomly
with equal probabilities from all pairs of genes in the current graph $s$.

The acceptance ratio of a proposed graph $s^\prime$
to $s$ can be calculated by
\begin{eqnarray}\label{eq-acceptrate-state}
\qquad\alpha_{G_t} & =& \frac{P(G_t=s^\prime\vert G_{-t}, D_{1\dvtx T},
\lambda)}{P(G_t=s\vert G_{-t}, D_{1\dvtx T}, \lambda)} \frac{P(s\vert
s^\prime)}{P(s^\prime\vert s)}
\nonumber\\
& =& \cases{
\displaystyle\frac{P(G_{t+1}\vert G_t=s^\prime, \lambda )}{P(G_{t+1}\vert G_t=s,\lambda)} \frac{P(D_t\vert G_t=s^\prime)}{P(D_t\vert G_t=s)},\qquad\mbox{if }t=1,
\vspace*{5pt}\cr
\displaystyle\frac{P(G_t=s^\prime\vert G_{t-1}, \lambda )}{P(G_t=s\vert G_{t-1},\lambda)} \frac{P(G_{t+1}\vert G_t=s^\prime,
\lambda)}{P(G_{t+1}\vert G_t=s,\lambda)} \frac{P(D_t\vert G_t=s^\prime)}{P(D_t\vert G_t=s)},
\vspace*{2pt}\cr
\hspace*{185pt}\mbox{if }t = 2,\ldots,T-1,
\vspace*{4pt}\cr
\displaystyle\frac{P(G_t=s^\prime\vert G_{t-1}, \lambda
)}{P(G_t=s\vert G_{t-1},\lambda)} \frac{P(D_t\vert G_t=s^\prime)}{P(D_t\vert G_t=s)},\qquad\mbox{if }t=T}\nonumber\\[-87pt]\\[47pt]\nonumber
\end{eqnarray}
in which the transition probability is
\begin{equation}
\label{eq-tr} P(G_t\vert G_{t-1}, \lambda) =
\frac{\lambda(1-\lambda)^{\varepsilon_t
-1}}{\sum_{\varepsilon_t^{\prime}=0}^{n_e}{n_e \choose
\varepsilon_t^{\prime}}\lambda(1-\lambda)^{\varepsilon_t^{\prime}-1}},
\end{equation}
where $n_e=n(n-1)$ is the number of all possible edges;
$\varepsilon_t=\llVert{A_t-A_{t-1}}\rrVert_1$, $A_{t}$~and~$A_{t-1}$ are the
adjacency matrices of network $G_{t}$ and $G_{t-1}$, respectively.
The normalising constant in equation~(\ref{eq-tr}) can be computed in advance,
as long as the number of vertices is fixed.

The proposal mechanism allows the sampler to transit from the current state
to any other state; the resulting Markov chain is therefore both
aperiodic and
irreducible. Moreover, the detailed balance condition is guaranteed by the
formulation of the Hastings ratio. Therefore, the proposed structural MH
algorithm is a correct MCMC sampler.\vspace*{9pt}

\textit{Sampling $\lambda$}.
The parameter $\lambda$ is sampled based on the Metropolis--Hastings
algorithm as
well. According to Bayes' theorem, the posterior probability of
$\lambda$ can
be computed as follows:
\begin{eqnarray}
P(\lambda\vert G_{1\dvtx T}, D_{1\dvtx T}) & \propto& P(D_{1\dvtx T},
G_{1\dvtx T}\vert\lambda) P(\lambda)
\nonumber\\[-4pt]\\[-15pt]
&=&\pi\prod_{t=2}^{T}P(G_t
\vert G_{t-1}, \lambda)\prod_{t=1}^{n}P(D_t
\vert G_t),\nonumber
\end{eqnarray}
where the equality statement assumes that the prior probability follows
a uniform distribution.

To constrain $\lambda$ between $0$ and $1$, we re-parameterise
$\lambda$ by the
sigmoid function, such that $\lambda= S(\kappa)$ where $S(\kappa) =
\frac{1}{e^{-\kappa}+1}$. Accordingly,\vspace*{1pt} the posterior probability of~$\kappa$ is
scaled by the determinant of the Jacobian (in this case, the Jacobian
is a~scalar):
\begin{eqnarray}
P(\kappa\vert G_{1\dvtx T}, D_{1\dvtx T}) & = &P(\lambda\vert
G_{1\dvtx T}, D_{1\dvtx T}) \frac{\partial
S(\kappa)}{\partial\kappa}
\nonumber\\[-8pt]\\[-8pt]
& = &P(\lambda\vert G_{1\dvtx T}, D_{1\dvtx T}) S(\kappa)
\bigl(1-S(\kappa) \bigr).\nonumber
\end{eqnarray}

\begin{algorithm}[b]
\caption{MCMC sampling algorithm for HM-NEMs}\label{alg1}
\begin{algorithmic}[1]
\Require$\mathcal{D}$, $\mathcal{G}^{(0)}$, $\lambda^{(0)}, \kappa
^{(0)} =
L(\lambda^{(0)}), \sigma$ \Comment{$L(\cdot)$ is the logit
function} \For{$i
= 1 \to N$} \For{$t = 1 \to T$} \State Propose $G_t^{*}$ by randomly
flipping an edge in $G_t^{(i-1)}$ \State Compute acceptance ratio
$\alpha_{G_t}^{(i)}$ with equation~(\ref{eq-acceptrate-state}) \State Draw
$u_{G_t}^{(i)}\sim\operatorname{Uniform}[0, 1]$ \If{$u_{G_t}^{(i)} <
\alpha_{G_t}^{(i)}$} \State Set $G_t^{(i)} = G_t^{*}$ \Else\State Set
$G_t^{(i)} = G_t^{(i-1)}$
\EndIf
\EndFor
\State Propose $\kappa^{*}$ from $\mathcal{N} (\kappa, \sigma)$
\State
Compute acceptance ratio $\alpha_{\kappa}^{(i)}$ with equation~(\ref
{eq-acceptrate-param}) \State Draw $u_{\kappa}^{(i)}\sim
\operatorname{Uniform}[0, 1]$ \If{$u_{\kappa}^{(i)} < \alpha_{\kappa
}^{(i)}$} \State
Set $\kappa^{(i)} = \kappa^{*}$ \Else\State Set $\kappa^{(i)} =
\kappa^{(i-1)}$
\EndIf
\State$\lambda^{(i)} = S(\kappa^{(i)})$ \Comment{$S(\cdot)$ is the sigmoid
function}
\EndFor
\end{algorithmic}
\end{algorithm}

Letting $\kappa^{\prime} \vert\kappa\sim\mathcal{N} (\kappa,
\sigma)$, the
acceptance ratio of proposed $\kappa^\prime$ to $\kappa$ is
\begin{eqnarray}
\label{eq-acceptrate-param} \alpha_{\kappa} &=& \frac{P(\kappa^\prime
\vert G_{1\dvtx T},
D_{1\dvtx T})}{P(\kappa\vert G_{1\dvtx T},
D_{1\dvtx T})}\nonumber
\\
&=&\frac{\prod_{t=2}^{n} P(G_t\vert G_{t-1},
\lambda^\prime)}{\prod_{t=2}^{n} P(G_t\vert G_{t-1}, \lambda)} \frac
{S(\kappa^\prime)(1-S(\kappa^\prime))}{S(\kappa)(1-S(\kappa))} \frac
{P(\kappa\vert\kappa^\prime)}{P(\kappa^\prime\vert\kappa
)}
\\
&=&\frac{\prod_{t=2}^{n} P(G_t\vert G_{t-1}, \lambda^\prime)}{\prod_{t=2}^{n}
P(G_t\vert G_{t-1}, \lambda)} \frac{S(\kappa^\prime)(1-S(\kappa^\prime
))}{S(\kappa)(1-S(\kappa))}.\nonumber
\end{eqnarray}

The complete sampling algorithm is described by the following pseudocode
(Algorithm~\ref{alg1}). 

\textit{Expected network}.
Let $\mathcal{A}:= \{A_{t}\}_1^T$ be the adjacency matrices of
$\mathcal{G}$. The expected time-varying network
$E[\mathcal{A}]=\{E[A_t]\}_1^T$ is computed by averaging over all adjacency
matrices of
$\mathcal{G}$ in the estimated posterior distribution obtained from
the sampling
result:
%
\begin{equation}
\label{expnet} E[A_t] = \sum_{A_t}
A_t P(A_t|D_{1\dvtx T}) = \frac{1}{N-N_b} \sum
_{i=1}^{N-N_b} A^{(i)}_t,
\end{equation}
where $N_b$ is the number of burn-in samples and $N$ is the total
number of
samples.

\section{Simulation studies}
\label{sim}
To evaluate the performance of the proposed MCMC sampling algorithm for HM-NEMs,
we conducted simulation studies on \textit{in silico} data generated from
artificially constructed networks. Each perturbation data set is
simulated by
the following steps:

\begin{enumerate}[4.]
\item For a given number of pathway components $n$,
for the first time frame we randomly flip 10\% off-diagonal entries of
a zero adjacency matrix to generate a~directed graph. This step is repeated
until the graph is transitively closed.\footnote{Here we generate
transitively closed graphs for the convenience to simulate data from
NEMs. HM-NEMs, however,
do not constrain the networks in the transitively closed graph space.}

\item For $t=2, 3, \ldots, T$, the network state at time $t$ is generated
by transiting
the previous state at time $t-1$. In detail,
we transit the previous network by flipping a random
number of off-diagonal entries $n_{er}$ following the distribution
${n_e \choose
n_{er}}\lambda(1-\lambda)^{n_{er}-1}$.
This step is also repeated until the graph generated for time frame $t$
is transitively closed.
%
\item For a network at each time frame, attach $n_r$ reporter genes to each
pathway component. For each component, generate perturbation data which
are 1s standing for
downstream effects and 0s for noneffects. The perturbation data are
duplicated for $n_p$ times to model biological replicates.
\item Add false negatives and false positives to the data generated in
the above
steps by randomly flipping $\alpha$ (\%) true negatives and
$\beta$ (\%) true positives.
\end{enumerate}

\subsection{Convergence diagnosis}\label{secconv}
To diagnose the convergence of proposed algorithm on networks that
transit states
rapidly, moderately and slowly, respectively, we did three simulations where
$\lambda$ was set to $0.1$, $0.5$ and $0.9$.
These values allow different dynamics in network evolving. Specifically,
they generate fast, moderate and slow varying networks, respectively.
The other parameters were fixed
to $n=6$, $n_{r}=4$, $T=8$, $n_{p}=3$, $\alpha=0.1$ and $\beta=0.1$.
It should be noted here that these parameters are set
according to a sophisticated biological application to embryonic stem cells,
where six genes were perturbed following phenotyping screening over
eight days.
For each simulation, a time varying network and corresponding
perturbation data
were generated according to the above protocol. Twenty independent runs
of MCMC
inference were performed on each data set over 52,000 iterations to
infer the
network and estimate $\lambda$. In principle, the posterior can always be
estimated with only one chain given that the chain is sufficiently
long. The
main purpose of using 20 runs in the simulation studies and the
following real
biological applications is to compute the Gelman and Rubin diagnostic to
inspect convergence. Furthermore, with multiple runs the posterior can be
estimated within a short time. The first 2000 samples of each run were treated
as the burn-in period.

\begin{figure}

\includegraphics{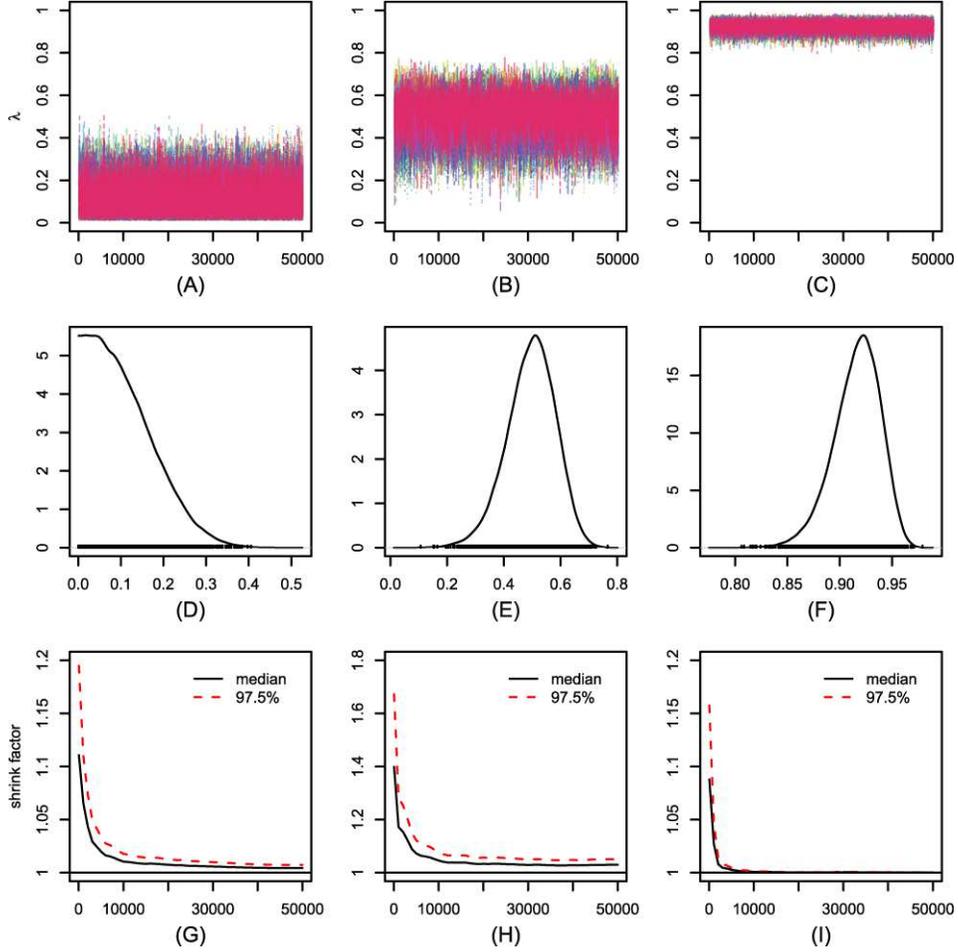}

\caption{Estimation and convergence diagnosis of $\lambda$.
\textup{(A)}, \textup{(B)} and \textup{(C)} are trace plots, \textup{(D)}, \textup{(E)} and \textup{(F)} are estimated
distributions, and \textup{(G)}, \textup{(H)} and \textup{(I)} illustrate logarithm $\sqrt{\hat{R}}$
statistics for $\lambda=0.1$, $\lambda=0.5$, $\lambda=0.9$, respectively.
Convergence is suggested when $\sqrt{\hat{R}}$ is close to~1.}\label{simlambda}
\end{figure}

We first investigate the performance of $\lambda$ estimation. To
achieve a good
rejection rate ($\sim$55\%) according to the efficient Metropolis
jumping rules
[\citet{gelman1996efficient}], we tuned $\sigma$, which is the
standard deviation
of the Gaussian proposal distribution (Table~\ref{ESSRR}). As shown in
Figure~\ref{simlambda}, the posterior distributions can be faithfully captured
by our sampling algorithm. As expected, the posterior means of estimated
$\lambda$ are very close to their corresponding true parameters
(Table~\ref{ESSRR}). To assess the efficiency of sampling, we also
compared the
effective sample sizes across three simulations (Table~\ref{ESSRR}).
To further evaluate the time of convergence to stationary
distributions, we
computed the $\sqrt{\hat{R}}$ statistic [proposed by
\citet{gelman1992inference}].
The time consumption (per run), using R on an Intel Xeon W3520 Quad-Core
2.67 GHZ with 8 GB RAM computer, does not show a big difference between these
three simulations (Table~\ref{ESSRR}).
We found that for all three simulations, our
algorithm converged within 52,000 iterations [Figure~\ref{simlambda}(G), (H) and (I)].
%
\begin{table}
\tabcolsep=0pt
\tablewidth=250pt
\caption{Performance of $\lambda$ estimation}\label{ESSRR}
\begin{tabular*}{\tablewidth}{@{\extracolsep{\fill}}@{}ld{4.2}d{3.2}d{3.2}@{}}
\hline
& \multicolumn{1}{c}{$\bolds{\lambda=0.1}$} & \multicolumn{1}{c}{$\bolds{\lambda=0.5}$} & \multicolumn{1}{c}{$\bolds{\lambda=0.9}$}\\
\hline
$\sigma$ & 2 & 0.65 & 0.65 \\
Posterior mean & 0.11 & 0.49 & 0.92 \\
Effective sample size & 4855 & 1301 & 4121 \\
Rejection rate & 0.53 & 0.53 & 0.54 \\
Time (sec) & 631.51 & 635.47 & 602.99 \\
\hline
\end{tabular*}
\end{table}

%

%
\begin{figure}

\includegraphics{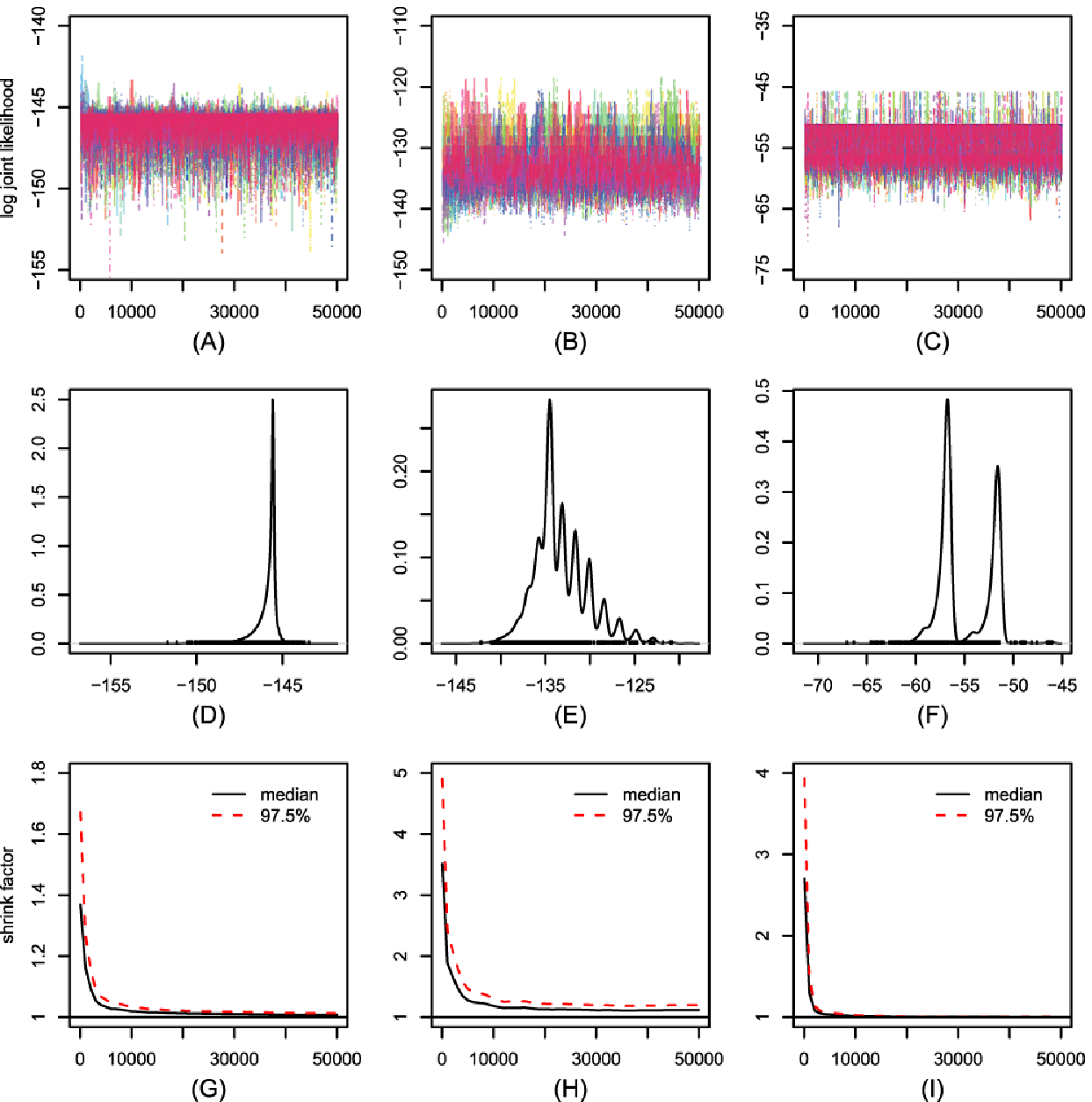}

\caption{Estimation and convergence diagnosis of log joint likelihood.
\textup{(A)}, \textup{(B)} and \textup{(C)} are trace plots, \textup{(D)}, \textup{(E)} and \textup{(F)} are
estimated distributions, and \textup{(G)}, \textup{(H)} and \textup{(I)} illustrate logarithm
$\sqrt{\hat{R}}$ statistics for $\lambda=0.1$, $\lambda=0.5$,
$\lambda=0.9$, respectively.}\label{simJLL}
\end{figure}

Visualising the posterior distribution of state path is difficult.
Instead, we evaluate the estimated distribution of the log joint likelihood.
As shown in Figure~\ref{simJLL}, multiple peaks appear in the
estimated log joint
likelihood distribution for $\lambda=0.9$, which indicates multiple optimal
state paths. Although in all three simulations the log joint likelihood
converge very quickly [as shown in Figure~\ref{simJLL}(G), (H) and
(I)], it
may become challenging for the algorithm to converge when there are
many more
optima.

\begin{figure}

\includegraphics{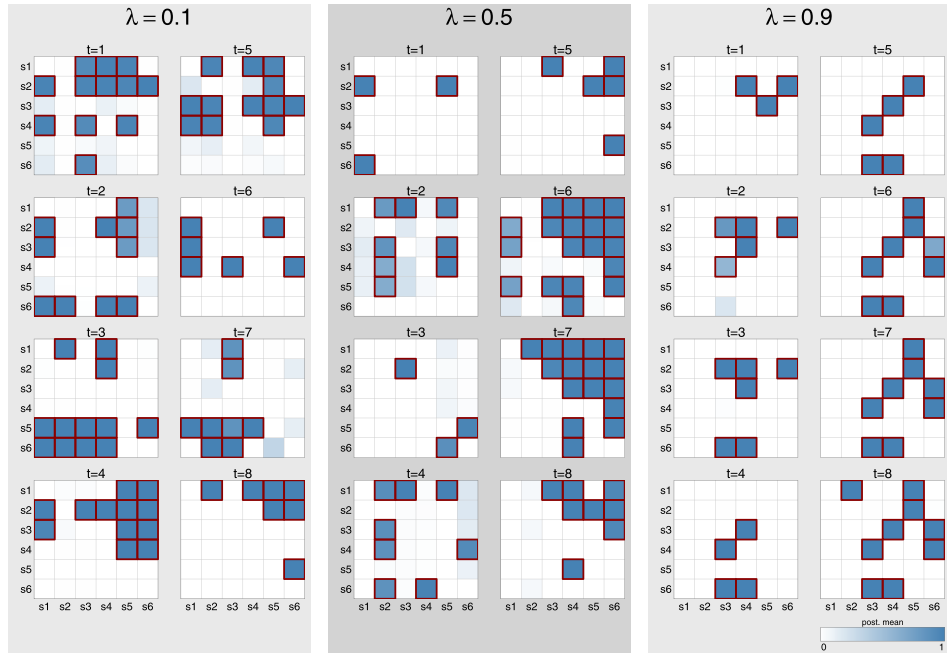}

\caption{Expected networks in simulation studies. The left, middle and
right panels correspond to $\lambda=0.1$, $0.5$ and $0.9$, simulating
networks that transit quickly, moderately and slowly, respectively.
Each heatmap illustrates the posterior means of edges directed from
pathway components in rows to their counterparts in columns.
The directed edges of true networks are outlined with bold borders in red.}\label{simheatmap}
\end{figure}

To evaluate the performance of network inference, we computed expected networks
using equation~(\ref{expnet}). Since here twenty independent runs were
performed for
each $\lambda$, the overall expected network is averaged over the expected
networks of all runs. The adjacency matrices of these overall expected networks
are illustrated by heatmaps in Figure~\ref{simheatmap}. Indeed, the inferred
networks seem to transit very dramatically for $\lambda=0.1$,
moderately for
$\lambda=0.5$ and very slowly for $\lambda=0.9$. Moreover,
considering 0.5 as a
cutoff to binarise these expected networks, we can compare to the true networks
generated. In all three simulations, our algorithm achieved 100\% high
sensitivities and 100\% specificities (Table~\ref{sensispeci}).\footnote{The
sensitivity, specificity and accuracy are computed by
$\mbox{TP}/(\mbox{TP}+\mbox{FN})$, $\mbox{TN}/(\mbox{TN}+\mbox
{FP})$ and
$\mbox{TP}+\mbox{TN}/(\mbox{TP}+\mbox{FN}+\mbox{TN}+\mbox{FP})$,
where TP,
TN, FP, FN are the number of true positive, true negative, false positive
and false negative directed edges, respectively.\label{ft1}}

Furthermore, we inspected the performance of HM-NEMs as a function of
the interval of time sampling with an additional simulation study
(details in
Appendix~\ref{app-timeSamp}). We found that a smaller time interval
tends to
give a better estimation of $\lambda$ and network (Figure~\ref{simtimesamp}).

In summary, the simulations from our model demonstrate the potential of
our sampling algorithm to infer evolving networks and ``smoothness''
under a
wide spectrum of network dynamics.

%

%
\begin{table}[b]
\tabcolsep=0pt
\tablewidth=250pt
\caption{Performance of network inference}\label{sensispeci}
\begin{tabular*}{\tablewidth}{@{\extracolsep{\fill}}@{}lccc@{}}
\hline
& $\bolds{\lambda=0.1}$ & $\bolds{\lambda=0.5}$ & $\bolds{\lambda=0.9}$ \\
\hline
Sensitivity & 100\% & 100\% & 100\% \\
Specificity & 100\% & 100\% & 100\% \\
\hline
\end{tabular*}
\end{table}
%

\subsection{Sensitivity analysis}
To assess the influence of several main factors, $\lambda$ and the error
probabilities $\alpha$ and $\beta$ to the performance of our algorithm,
we vary $\alpha$ and $\beta$ from $0.1$ to $0.5$ to simulate
artificial data sets
for the three networks (corresponding to $\lambda=0.1$, $0.5$ and $0.9$,
resp.) we have already generated in Section~\ref{secconv}.
For each combination of parameters ($\lambda$, $\alpha$ and $\beta
$), we
simulated 1000 random data sets and performed MCMC sampling over 12,000
iterations, which are sufficient for the algorithm to converge
according to
Figures~\ref{simlambda} and \ref{simJLL}, with the first 2000 as
the burn-in
period.
The mean Monte Carlo errors of estimated $\lambda$ and the accuracies
(footnote~\ref{ft1} in Section~\ref{secconv}) of expected networks
binarised with a cutoff of 0.5 were calculated for each parameter setting.

%
\begin{figure}

\includegraphics{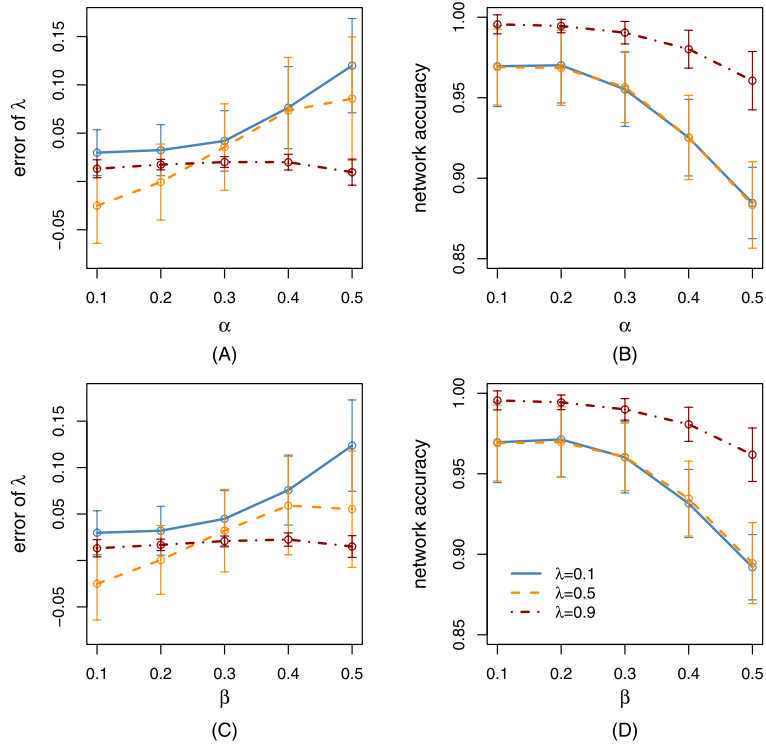}

\caption{Mean Monte Carlo errors of estimated $\lambda$ [(\textup{A}) and (\textup{C})] and
accuracies
of inferred expected networks [(\textup{B}) and (\textup{D})] as a function of $\alpha$
(when $\beta=0.1$) and
$\beta$ (when $\alpha=0.1$).
Lines colored in blue, orange and red correspond to $\lambda=0.1$, $0.5$
and $0.9$, respectively.
Error bars indicate one standard deviation of the mean.}\label{simerror}
\end{figure}

As shown in Figure~\ref{simerror}, $\alpha$ and $\beta$ affect more
the performance on networks that transit relatively fast
($\lambda=0.1$ and $0.5$) than those that transit slowly ($\lambda=0.9$),
indicating that our algorithm is more robust to data noise in smooth networks.
Nevertheless, no matter how $\lambda$ varies, our algorithm can still achieve
very promising performance (mean Monte Carlo error of $\lambda< 0.05$
and accuracy of inferred network $>$0.95) when $\alpha$ and
$\beta$ are $\leq30$\%, which is easily satisfied in real biological
experiments.

We further used ANOVA to quantify the relative contributions of $\alpha$,
$\beta$ and $\lambda$ to the performance of our algorithm.
Table~\ref{anovaA} combines the results of a design-based ANOVA of the
mean Monte Carlo error of estimated $\lambda$ and the accuracy of
inferred networks.
As expected, all factors and their interactions are very significantly
associated with performance ($p < 0.0001$).
Interestingly, $\lambda$ itself explains 30.41\% total variation of
the mean
Monte Carlo error of estimated $\lambda$, which is much more than the
8.21\% and
8.54\% explained by $\alpha$ and $\beta$, indicating that $\lambda$
estimation is more sensitive to the smoothness of networks.
However, network inference is more affected by noise in the perturbation
data, as $\alpha$, $\beta$ and their interaction account for 91.35\% total
variation.

%
\begin{table}
\tabcolsep=0pt
\caption{Analysis of variance of mean Monte Carlo errors of estimated
$\lambda$ and accuracies of inferred networks. All factors
and interactions were statistically significant with $p < 0.0001$}\label{anovaA}
{\fontsize{9}{10.7}\selectfont{\begin{tabular*}{\tablewidth}{@{\extracolsep{\fill}}ld{5.0} cd{6.2} d{3.2}d{6.2}@{}}
\hline
\multicolumn{2}{c}{} & \multicolumn{2}{c}{\textbf{Mean error of} $\bolds{\lambda}$} &\multicolumn{2}{c}{\textbf{Network accuracy}}\\[-6pt]
\multicolumn{2}{c}{} & \multicolumn{2}{c}{\hrulefill} &\multicolumn{2}{c}{\hrulefill}\\
& \multicolumn{1}{c}{\textbf{Df}} & \multicolumn{1}{c}{\textbf{Sum Sq}} & \multicolumn{1}{c}{\textbf{$\bolds{F}$-value}} & \multicolumn{1}{c}{\textbf{Sum Sq}} & \multicolumn{1}{c}{\textbf{$\bolds{F}$-value}}\\
\hline
$\lambda$ & 2 & 762.32 & 246{,}560.81 & 41.54 & 37{,}650.57 \\
$\alpha$ & 4 & 205.90 & 33{,}296.87 & 447.53 & 202{,}801.65 \\
$\beta$ & 4 & 214.02 & 34{,}609.80 & 425.06 & 192{,}617.84 \\
$\lambda\dvtx \alpha$ & 8 & 365.01 & 29{,}514.12 & 1.23 & 278.97 \\
$\lambda\dvtx \beta$ & 8 & 361.83 & 29{,}257.05 & 0.80 & 181.54 \\
$\alpha\dvtx \beta$ & 16 & 331.15 & 13{,}388.17 & 138.11 & 15{,}645.82 \\
$\lambda\dvtx \alpha\dvtx \beta$ & 32 & 150.69 & 3046.12 & 10.76 & 609.39 \\
Residuals & 74{,}925 & 115.83 & & 41.34 & \\
\hline
\end{tabular*}}}%
\end{table}\vspace*{-3pt}

Taken together, these sensitivity analysis results demonstrate that our sampling
algorithm can be reliably employed to infer evolving networks and estimate
the smoothness for perturbation data that are not extremely noisy.


\subsection{Coverage analysis}
We next investigated the frequentist coverage of Bayesian confidence intervals
as a function of $\alpha$ and $\beta$ in different contexts of
network transition.
For each combination of parameter $\alpha$ (0.1 to 0.5), $\beta$ (0.1
to 0.5) and
$\lambda$ (0.1, 0.5 and 0.9), we computed highest posterior density
(HPD) intervals
with 95\% nominal coverage probability for estimated $\lambda$ and log
joint likelihood.
The ``actual'' coverage probability was subsequently computed by the
proportion of the time
that the HPD interval contains the true $\lambda$ or log joint likelihood.

%
\begin{figure}

\includegraphics{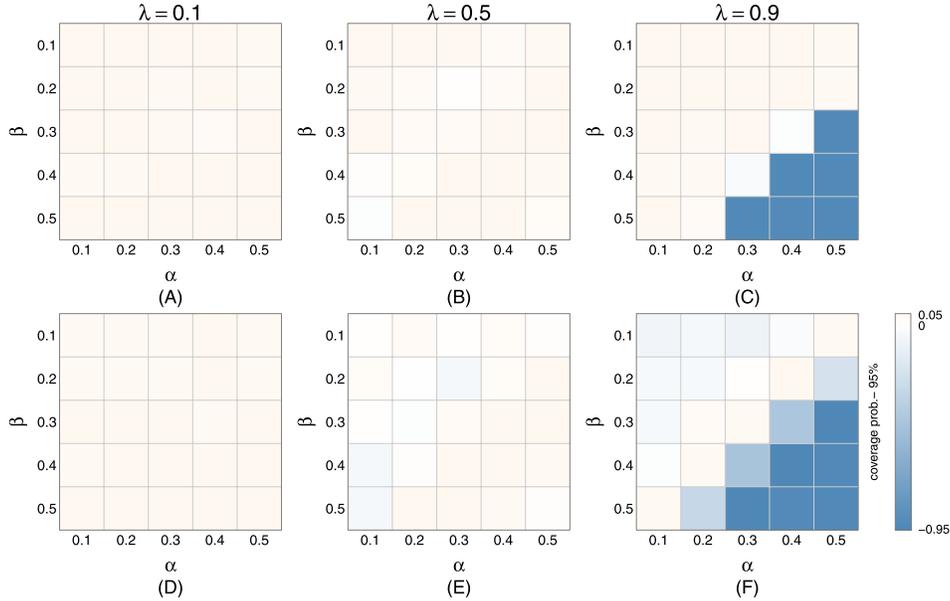}

\caption{Coverage as a function of $\alpha$ and $\beta$.
\textup{(A)}, \textup{(B)} and \textup{(C)} are heatmaps illustrating the difference between coverage
probability and 95\% nominal coverage probability of estimated $\lambda$
across different $\alpha$, $\beta$ and $\lambda$.
Similarly, \textup{(D)}, \textup{(E)} and \textup{(F)} correspond to coverage of the log joint likelihood.}\label{simcoverage}
\end{figure}

As illustrated in Figure~\ref{simcoverage}, the ``actual'' coverage probability
faithfully matches the nominal coverage probability across all $\alpha
$ and
$\beta$ when $\lambda=0.1$ and $0.5$.
When $\lambda=0.9$, the true $\lambda$ and log joint likelihood are
both outside
of the Bayesian intervals when $\alpha$ and $\beta$ are both
unreasonably high
($\ge$0.4).
However, even when $\alpha$ or $\beta$ are as high as $0.3$, we still
observed very good
coverage for $\lambda=0.9$ (Figure~\ref{simintervals} in
Appendix~\ref{app}).

Taken together, the coverage analysis results demonstrate that our algorithm
provides good coverage performance as long as the quality of perturbation
data is not extremely bad ($\alpha\leq0.3$ and $\beta\leq0.3$),
which is
often satisfied in real biological experiments.\vspace*{-2pt}

\section{Applications}\vspace*{-2pt}
\subsection{Application to neutrophil polarisation}\label{neutro}
Neutrophils are phagocytic immune cells that can detect and kill
bacteria very
quickly. Underlying the rapid\vadjust{\goodbreak} response of neutrophils to
chemoattractants is
the neutrophil polarity network, which upon stimulation progresses
through three
phases: instantaneous initiation, 2--3~min development and 10 min maintenance
before adapting. Three spatially and molecularly distinct cytoskeletal
modules---front (F), back (B) and microtubule (M) modules---have been
implicated to
be involved in the neutrophil polarity network [\citet{small2002microtubules}].
Despite various interactions identified between these three modules,
how they
crosstalk dynamically to regulate polarisation is still poorly understood.

To gain mechanistic insights to the interactions between the front,
back and
microtubule modules over time upon stimulation, \citet
{ku2012network} conducted systematic
pharmacological perturbations to the three modules and employed a
microscopy-based approach to quantify neutrophil polarisation phenotypes.
In detail, each module was targeted by two opposing
mechanistically distinct drugs: LasA (inhibitor) and Jas (enhancer) for
the F
module, Y27632 (inhibitor) and Calp (enhancer) for the B module, Noco
(inhibitor) and Taxol (enhancer) for the \mbox{M module}. Each perturbation experiment
was done with 2 to 6 replicates over 600 seconds after stimulation of
f-Met-Leu-Phe (fMLP), which is a strong chemoattractant directing
movements of
neutrophils towards bacteria. As a control, responses of neutrophils to fMLP
without drug treatment were also investigated in 20 replicates. The
perturbation effect of the three modules was monitored by three protein
markers: F-actin for the F module, $\alpha$-tubulin for the M module, and
p-MLC2 for the B module, respectively.
For each protein marker, intensity and polarity were quantified based
on image
analysis.
In this application the pathway components are the three perturbed
modules (F, B and M), while effect reporters are the three biochemical
markers (F-actin, p-MLC2 and $\alpha$-tubulin).

In the perturbation experiments, cells were fixed at 11 nonuniform time points
from 0 to 600 seconds. The original time series were interpolated and smoothed
to generate a response curve for each replicate, perturbation, protein marker
and phenotype [details in \citet{ku2012network}]. Without loss of generality,
here we focus on the interpolated polarisation response data of
perturbation by enhancers of Jas, Calp and Taxol across 41 time
points with the same sampling interval of 15 seconds.

At time point $t$ for perturbation $k$, a $t$-score was
computed by
comparing the observed phenotype $i$ with the reference distribution of
phenotypes
in control experiments. The $t$-value was used to compute the
probability ($p_{ikt}$) that the perturbation phenotype is different from
controls based on one sample Bayesian $t$-test [\citet{rouder2009bayesian}].
As the relationships between the three modules (F, B and M) and the effect
reporters (F-actin, p-MLC2 and $\alpha$-tubulin) are already known, we use
equation~(\ref{eq1b}) to calculate the emission probability $P(D_t \vert
G_t)$, which
can be written as $P(d_{ikt}\vert G_t)=p_{ikt}$ if $S_{jkt}=1$ and
$P(d_{ikt}\vert G_t)=1-p_{ikt}$ if $S_{jkt}=0$. Having obtained the
probabilities for all phenotypes, perturbations and time points, we applied
HM-NEMs to infer the dynamic interplay between the F, B and M modules.

%
\begin{figure}
\includegraphics{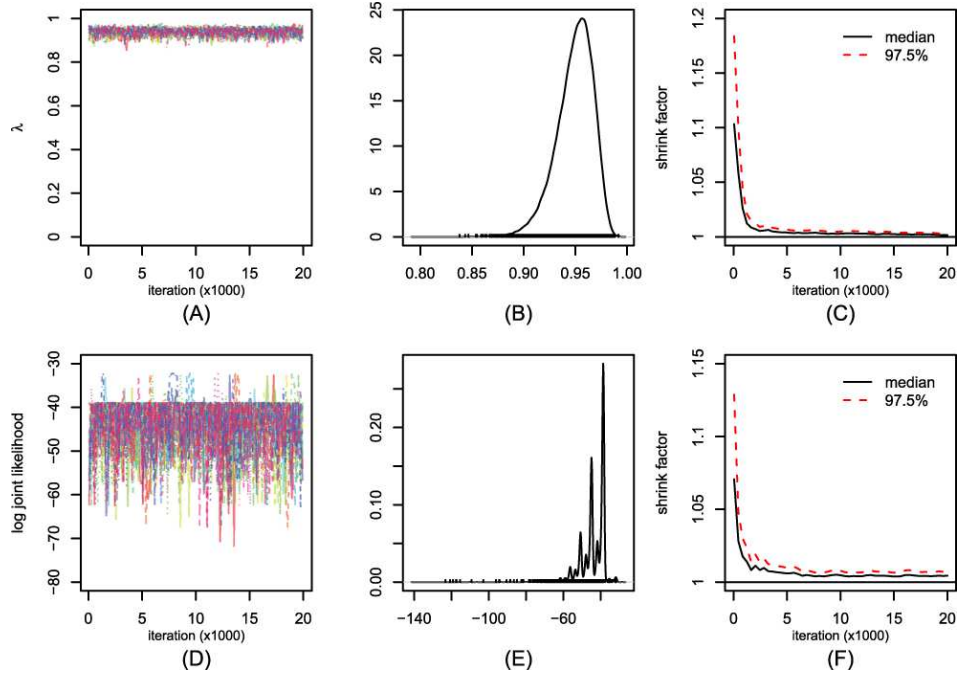}
\caption{Parameter estimation and convergence diagnosis in the
application to
neutrophil polarisation. \textup{(A)}, \textup{(B)} and \textup{(C)} are trace plot, posterior
distribution and logarithm $\sqrt{\hat{R}}$ statistics for $\lambda$
estimation. \textup{(D)}, \textup{(E)} and \textup{(F)} are trace plot, posterior distribution and
logarithm $\sqrt{\hat{R}}$ statistics for the log joint likelihood estimation.}\label{neutrophilconv}
\end{figure}

Twenty parallel runs of MCMC sampling were performed to do parameter estimation
and network inference. Each sampling was run for 22,000 iterations, and the
first 2000 were considered as the burn-in period. As shown in
Figure~\ref{neutrophilconv}, we observed a very fast convergence for both
$\lambda$ and the log joint likelihood. The posterior mean of $\lambda
$ is
$\sim$0.95, indicating that the polarity network of neutrophils
progresses very
smoothly.

%
\begin{figure}

\includegraphics{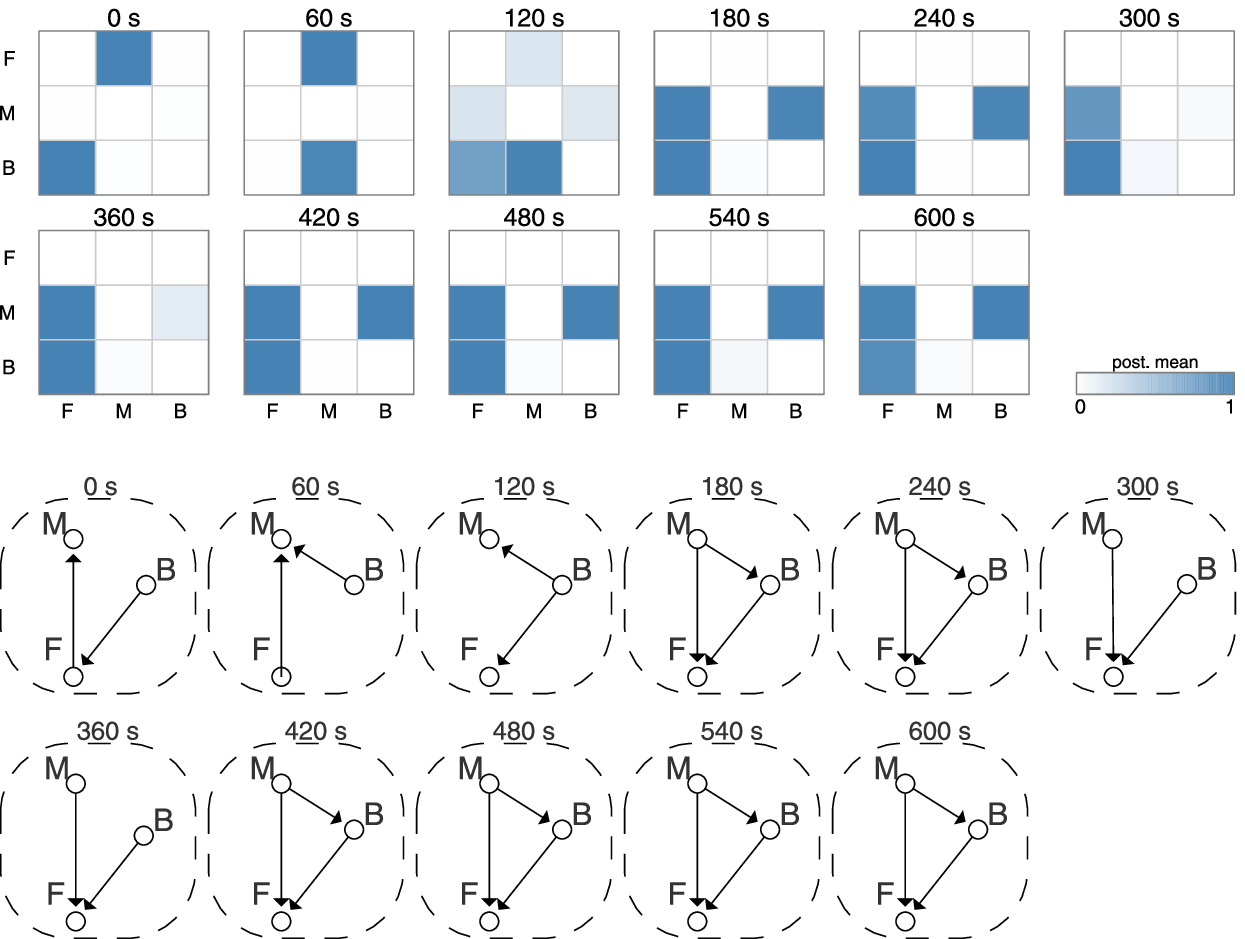}

\caption{Expected (upper panel) and discretised (lower panel) neutrophil
polarisation network. The evolving network identified by HM-NEMs precisely
captures the three-phase progression of the polarity network of neutrophils:
polarisation initiation (0 sec), polarisation development (60 to 120~sec)
and polarisation maintenance (180 to 600~sec). The discretised network
was produced by R package RedeR~[\citet{castro2012reder}].}\label{neutrophilgraph}
\end{figure}

The overall expected network was summarised over all twenty parallel
chains, and
the snapshots at every 60 seconds are
illustrated in the heatmaps of Figure~\ref{neutrophilgraph}. The vast
majority of the posterior means of pairwise interactions are either
close to 1
or 0, suggesting that the signalling interactions can be identified by our
algorithm without ambiguity. The overall expected network was further binarised
based on a cutoff of posterior mean at 0.5 (Figure~\ref{neutrophilgraph}).
Intriguingly, the feedforward signalling of M module to B and F modules
dominates the maintenance phase ($180$ to $600$ sec), which is consistent
with the persistant crosstalks identified in~\citet
{ku2012network} using a $z$-score-based
approach. Signalling interactions inferred at early stages
($0\sim120$ sec), mainly from the back and front modules to the microtubule
module, coincide with the transient crosstalks, which are known to happen
during the initiation and development phases of neutrophil
polarisation. Taken
together, the evolving network inferred by HM-NEMs captures
dynamic crosstalks between the front, back and microtubule modules
underlying neutrophil polarisation.

\subsection{Application to mouse embryonic stem cells}
\label{esc}
There has been a wealth of studies on the self-renewal and differentiation
mechanisms of embryonic stem cells (ESCs) for decades. \citet
{ivanova2006dissecting}
conducted an integrated approach to reveal potential regulators
participating in the self-renewal process of murine ES cells. They
first used
shRNAs (short hairpin RNAs) to knock down potential transcription
factors and
identified six regulators (\textit{Nanog}, \textit{Oct4}, \textit{Sox2},
\textit{Esrrb}, \textit{Tbx3} and \textit{Tcl1}). Transcriptome
dynamics were
subsequently monitored after perturbation of each of these six factors over
eight days using microarrays. Using a simple clustering analysis, they found
that the six regulators are involved in two global pathways regulating ESC
self-renewal~[\citet{ivanova2006dissecting}]. However, how these identified
factors interact with each other remains unclear. Here, we attempt to
address this challenge by HM-NEMs.

\begin{figure}

\includegraphics{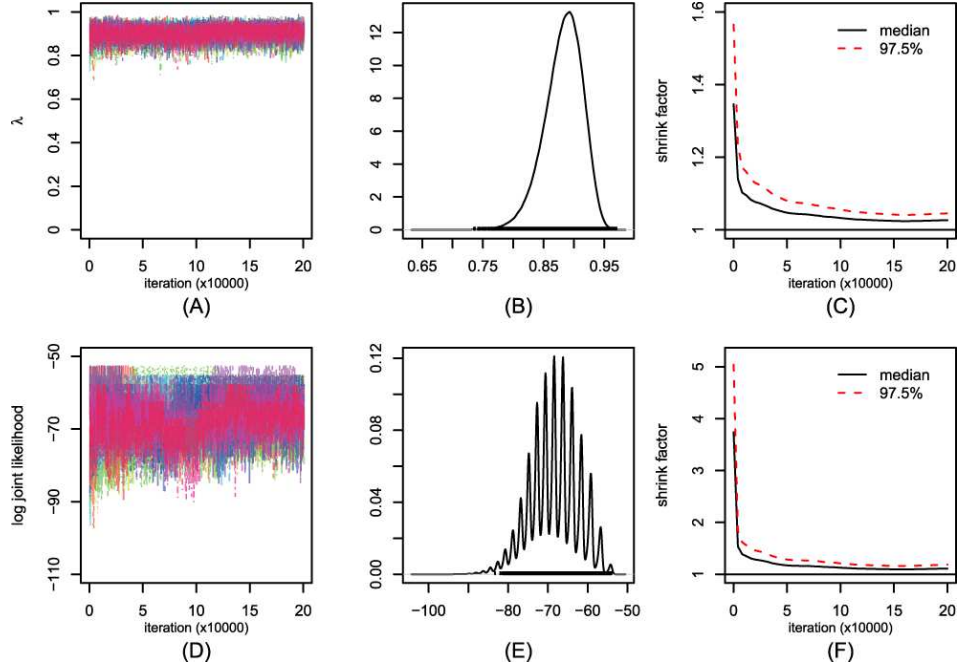}

\caption{Parameter estimation and convergence diagnosis in the
application to
mouse embryonic stem cells. \textup{(A)}, \textup{(B)} and \textup{(C)} are trace plot, posterior
distribution and logarithm $\sqrt{\hat{R}}$ statistics for $\lambda$
estimation. \textup{(D)}, \textup{(E)} and \textup{(F)} are trace plot, posterior distribution and
logarithm $\sqrt{\hat{R}}$ statistics for the log joint likelihood
estimation.}\label{escconv}
\end{figure}

The raw gene expression data were preprocessed and discretised
following the
same strategy in~\citet{anchang2009modeling}. From the discretised perturbation
data, we applied HM-NEMs to reconstruct the ESC self-renewal network. Twenty
independent runs of MCMC sampling were performed over 202,000 iterations
including the first 2000 burn-in period.
To speed up convergence, the starting network at each time point was
inferred by static NEMs with a greedy hill-climbing algorithm
[\citet{markowetz2006phd}].
As shown in Figure~\ref{escconv}, the
Markov chains of $\lambda$ and log joint likelihood converge very quickly.
Similar to the neutrophil polarisation network, the posterior mean of estimated
$\lambda$ is $\sim$0.88, suggesting that the network transition
underlying ESC
early differentiation is also very smooth.

%

%
\begin{figure}

\includegraphics{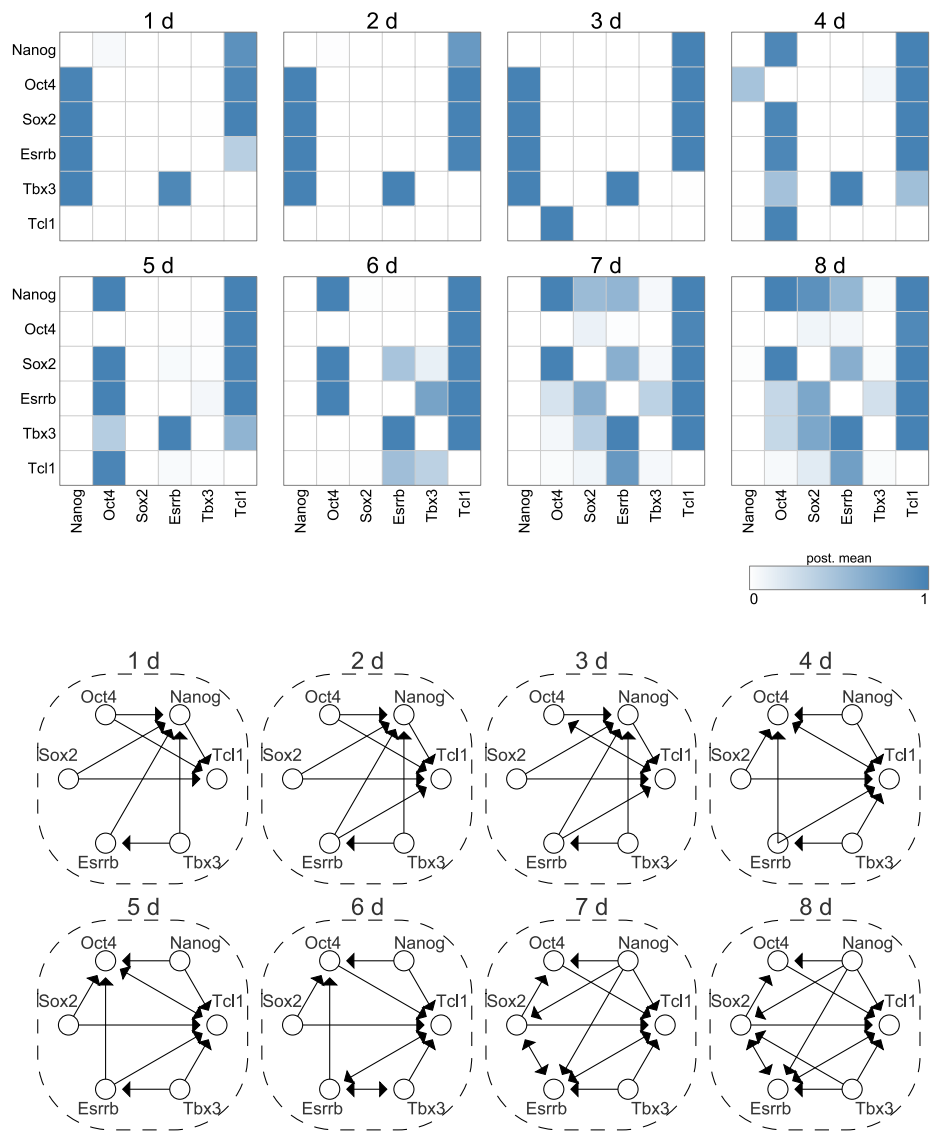}

\caption{Expected (upper panel) and discretised (lower panel) network
in the
application to mouse embryonic stem cells. The network identified by
HM-NEMS suggests that the feedback regulations between Nanog, \textit{Sox2} and \textit{Oct4}
may underlie early differentiation of mouse ESCs.}\label{escgraph}
\end{figure}

We computed the overall expected network across all twenty chains, and observed
that the vast majority of signalling interactions between the six
regulators are
quite deterministic (illustrated by the heatmaps in Figure~\ref{escgraph}).
Using a cutoff of posterior mean at 0.5, we further discretised the overall
expected network (Figure~\ref{escgraph}). Interestingly, the network suggests
both feed-forward and feedback regulations during early differentiation
of ESCs.
One feed-forward loop (day 4 to day~8) is found between \textit{Sox2},
\textit{Oct4} and \textit{Tcl1}, which can be validated by the transcriptional
regulations of \textit{Sox2} on \textit{Oct4}~[\citet{masui2007pluripotency}] and
\textit{Oct4} on \textit{Tcl1}~[\citet{matoba2006dissecting}]. Another
feed-forward loop (day 4 to day 8) is between \textit{Tbx3}, \textit{Esrrb} and
\textit{Tcl1}, which is less characterised in the literature. However, the
network constituted by them is found to be critical to block the differentiation
into epiblast-derived lineages~[\citet{ivanova2006dissecting}]. These two
feed-forward loops are mainly regulating the expression of \textit{Tcl1}, which
was shown
to be important for ESC proliferation but not differentiation~[\citet{matoba2006dissecting}, \citet{ivanova2006dissecting}].

Feedback interactions are mainly found between \textit{Nanog} and
\textit{Oct4}/\textit{Sox2} and \textit{Tbx3}/\textit{Esrrb}.
\textit{Nanog} is
found downstream of \textit{Sox2}/\textit{Oct4} during the early
stage (day 1 to
day 3), but upstream of \textit{Oct4} and/or \textit{Sox2} after day
4. The
feedback regulations between \textit{Nanog} and \textit{Sox2}/\textit
{Oct4} are
known to be critical for maintaining the pluripotency of ESCs~[\citet{loh2006oct4}].
\textit{Nanog} can bind to the promoter regions of
\textit{Oct4} and \textit{Sox2}, while the \textit{Oct4}--\textit{Sox2}
heterodimer can also bind to the promoter region of \textit
{Nanog}~[\citet{boyer2005core}, \citet{loh2006oct4}].
Thus, the feedback between \textit{Nanog}
and \textit{Sox2}/\textit{Oct4} may occur at the transcription level. The
feedback regulations between \textit{Nanog} and \textit{Tbx3}/\textit
{Esrrb} are
also implicated in the literature. \textit{Nanog} is known to positively
regulate expression of \textit{Esrrb}, while \textit{Tbx3} and
\textit{Esrrb}
can also enhance \textit{Nanog} expression~[\citet{loh2006oct4},
\citet{van2008estrogen}, \citet{niwa2009parallel}].

Taken together, we hypothesize that the time-varying signalling network
inferred by HM-NEMs underlying the early differentiation of ESCs may involve
two stages. During the early stage (day 1 to day~3), \textit{Sox2} and
\textit{Oct4}
positively regulate expression of \textit{Nanog} so that ESCs maintain its
self-renewal. During the late stage (after day 4), \textit{Nanog}
starts to
regulate \textit{Oct4} and/or \textit{Sox2} and lead ESCs to differentiate.
Our hypothesis can be confirmed in part by previous findings that high
\textit{Nanog} expression is important for ESCs to possess high self-renewal
efficiency, whereas low \textit{Nanog} expression is associated with increased
differentiation propensity~[\citet{kalmar2009regulated}, \citet{navarro2012oct4}].
Nonetheless, \mbox{biological} experiments should be conducted to further
validate the
reconstructed network.

\section{Discussion}
In this paper we propose hidden Markov nested effects models for reconstructing
signalling networks evolving over time.
We developed a MCMC sampling algorithm to infer the most probable state
path (the evolving network) while estimating the parameter $\lambda$
that indicates
the intrinsic feature of network evolutions.
With simulations from the model, the proposed
MCMC sampling algorithm was shown to work efficiently on networks under
a wide
spectrum of network dynamics as long as the perturbation data is not
extremely noisy.
We also demonstrated the model's potential to infer evolving networks underlying
dynamic biological processes by two real applications.\vspace*{9pt}

\begin{figure}

\includegraphics{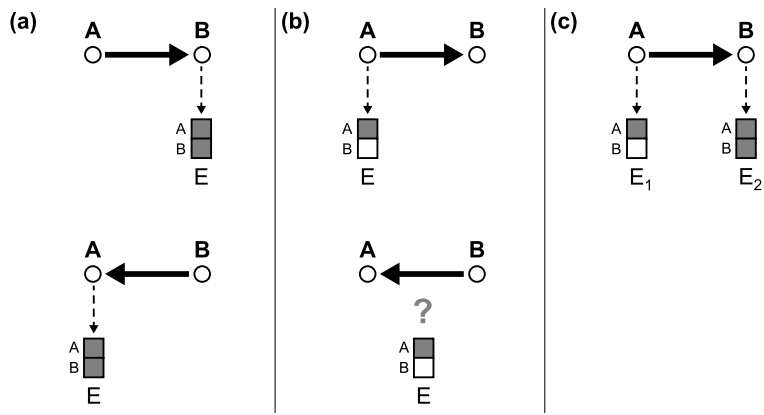}

\caption{NEM examples involving only two pathway components A and B.
Inference success depends on the information in the data (boxes for
effect reporter status under perturbation of A or B: shaded${}={}$effect;
white${}={}$no effect). Dashed edges show error-free (and thus most likely)
effect reporter position given the A--B structure and the data. \textup{(a)}
Single effect reporter showing effects under both perturbations makes
structures indistinguishable; \textup{(b)}~single effect reporter showing effect
under only one perturbation (in this case A) prefers structure where
the perturbed gene is on top; \textup{(c)}~generally we have more than one
effect reporter and this is an example of a small subset structure on
two reporters.}\label{singleEgene}
\end{figure}

\textit{Identifiability}.
Inference of NEMs is based on the model posterior, which combines the
prior distributions on pathway structure and positions of effect
reporters with the data on effect reporter states under perturbations.
We generally choose uniform priors and, for example, the position of
each effect reporter is equally likely at each pathway component. In
the following, we use a toy example involving only two pathway
components (Figure~\ref{singleEgene}) to discuss the identifiability
of NEMs. We will show that whether the two structures in the example
($A\rightarrow B$ and $B\rightarrow A$) are distinguishable or not
depends on the data observed at the effect reporter E.

%

If a single effect reporter shows effects under both perturbations
[Figure~\ref{singleEgene}(a); a special case of two perturbations
showing the same profile over all effect reporters], the structures
$A\rightarrow B$ and $B\rightarrow A$ are indeed indistinguishable. In
this symmetric case the effect reporter can always be attached to the
downstream gene without preferring one structure to the other.
Generally, NEMs model subset relations and in this situation the set of
effects after perturbing~A ($=\{E\}$) is identical to the set of
effects after perturbing~B ($=\{E\}$), which indeed offers no
information on how to order A and B. This issue has been identified
already in the first NEM papers~[\citet{markowetz2005non}, \citet{markowetz2007nested}]
and it is the reason why we often use
bidirectional arrows to indicate pairs of pathway components with (up
to noise) identical effect profiles. For larger networks we generally
merge all these nodes into a joint node and, as a result, infer a
hierarchy of clusters of genes, instead of a hierarchy of individual
genes~[\citet{markowetz2007nested}].

In practice, however, this identifiability problem may not be dramatic,
because effect profiles are generally not identical, especially with
gene expression readout on thousands of genes like in our second case study.
As the simplest example,\vadjust{\goodbreak} imagine that the effect reporter only shows an
effect under one perturbation (say~A) but not the other [Figure~\ref{singleEgene}(b)]. Even with only a single reporter, this small change in
the data improves the situation drastically: now the set of effects
after perturbing~A ($=\{E\}$) is a superset of the set of effects after
perturbing~B ($=\varnothing$). Our likelihood (and also the marginal
likelihood averaging over all reporter positions) now prefers the
$A\rightarrow B$ structure, because it allows to attach E to A without
incurring any false positive or negative effects, while no error-free
attachment is possible for $B\rightarrow A$ (if we attached it to B,
the reporter profile would be completely wrong; and if we attached it
to A, the effect of perturbing~B would still be missing).

In real applications there are usually more effect reporters than
pathway components. Our scenario is thus better represented by a graph
with two effect reporters [Figure~\ref{singleEgene}(c)]. If both
reporters show effects under both perturbations, we are back in the
situation of Figure~\ref{singleEgene}(a), but generally we observe
patterns like the one shown here: the set of effects after perturbing~A
($=\{E_1,E_2\}$) is a superset of the effects of perturbing~B ($=\{E_2\}
$), which allows unique identification of the A--B structure and
reporter assignment.

Theoretically, it has also been proved that NEMs are identifiable for
sufficiently `good' data, which can be satisfied by a sufficient number
of replicate measurements, and the pathway graph and the assignment of
reporters to pathway components are unique up to reversals [see
Theorems 1~and~3 in \citet{Tresch2008}].

In HM-NEMs, the identifiability of hidden states (in our case networks,
also under given parameters) is expected to be at least the same as
estimating the states separately from different time points with a
static model. One could see a \mbox{HM-NEM} as a static NEM with an
informative prior estimated from different time points. In this regard,
it may be easier to estimate a network at a given time point with
HM-NEM than a static NEM.\vspace*{9pt}

\textit{Label switching}.
In HMMs, the label switching problem arises from the fact that the
observation model is the same as the generative process in a mixture
model. As the labels for each component are exchangeable, the marginal
likelihood (which is the objective function for parameter estimation)
after integrating out the unknown labels is invariant to permutations
of parameters. When the hidden states are the parameter labels, HMMs
can be seen as mixture models with a Markov prior over the labels. The
marginal likelihood is again invariant to permutations of the
parameters. Thus, the label switching got introduced into HMMs.

However, when the hidden states are the parameters of the distribution
rather than the labels of the parameters, the situation changes. For
the observation model there is only one parameter rather than $K$
number of parameters in the marginal likelihood. Let us illustrate this
with a 2-state Gaussian HMM. Assuming the same covariance matrix, at a
give time point, the observation model is $N(\mu_k, \Sigma)$, $k \in
(1, 2)$. Therefore, there are two unknown parameters $\mu_1$ and $\mu
_2$ after integrating out the unknown labels. Now consider for the same
observed data, but the hidden variable is $\mu$ itself. Integrating
out $\mu$, there is only $\Sigma$ left in the marginal likelihood.
Hence, there is nothing to switch any more.
In HM-NEMs, the hidden states are the network topologies which are the
parameters of NEMs rather than labels. Therefore, we think that label
switching may not be a problem for \mbox{HM-NEMs}.\vspace*{9pt}

\textit{Generality}.
Although \mbox{HM-NEMs} belong to the family of structural HMMs, it is specifically
tailored for indirect data from systematic perturbation screens.
\mbox{HM-NEMs} extend
classical nested effects models which infer static signalling networks.
With an
identity transition kernel ($\lambda\rightarrow1$), HM-NEMs are
identical to
static NEMs since the graph chosen at the first time frame persists
till the
end.
On the other hand, when $\lambda\rightarrow0$, there is no structural
dependencies between consecutive time points (the process loses its
memory), and
independent NEMs are fitted for each observation time point. To strike the
balance between a static view of the data and the negligence of time dependence,
HM-NEMs use a transition kernel to model the nature of biological
networks that
transit smoothly over time.
Beyond the NEMs family, there are also generalisable elements in the
model. Particularly, the transition
probability setting could be applied to other structural HMM including DBNs.\vspace*{9pt}

\textit{Scalability}.
In HM-NEMs, the cardinality of the state space grows exponentially with
the size
of the signalling network.
The traditional Baum--Welch algorithm for HMM suffers from the time
complexity of $O(K^2TM)$, where $M$ is the number of the EM iterations,
$K=2^{n_s(n_s-1)}$ is the size of network state space for~$n_s$ pathway
components and $T$
is the number of time points. As the networks grow in size, the use of
this type
of method might be prohibited. Such a scalability problem makes MCMC
particularly
appealing to HM-NEMs, since the Monte Carlo estimator does not suffer
the cure of
dimensionality. However, a large state space may potentially lead to highly
multimodal posterior or low identifiability, both of which can result
in poorly
mixing Markov chains.\vspace*{9pt}

\textit{More efficient inference methods}.
Both scaling and generalising HM-NEMs demand efficient inference strategies.
Our inference method is an elementary single-site update Gibbs sampler. The
sampler could be easily trapped in a local region of the posterior due
to the
fact that the hidden variables are updated sequentially. This could be
solved by
employing block-type update schemes which sample a part of or even the entire
state path in one go. In addition, a significant efficiency gain could be
obtained by sampling $\lambda$ with the recently developed manifold MCMC
approaches~[\citet{girolami2011riemann}].
Another improvable point is to use a better structural
MCMC which efficiently explores the network topologies. To this end,
one could
convert the network structure into ordered space~[\citet{friedman2003being}]. Other
advances in structural MCMC in DAGs, including those in~\citet{grzegorczyk2008improving},
could be applied in our setting as well. These are the algorithmic
avenues we
are currently pursuing.\vspace*{9pt}

\textit{Biological applicability and implications}.
In this paper we also demonstrated the potential of HM-NEMs to gain biological
insights to the network transition underlying complex dynamic biological
processes. In the application to neutrophils, HM-NEMs capture the transition
between initiation, development and maintenance phases during neutrophil
polarisation. In another application to mouse \mbox{embryonic} stem cells
(ESCs), the
time-varying network inferred by HM-NEMs suggests that underlying early
differentiation of ESCs may be the feedback regulations between \textit{Nanog},
\textit{Sox2} and \textit{Oct4}. Our results on these two real biological
applications are in part consistent with recent findings in the literature,
and generate an intriguing hypothesis about the mechanisms of network
evolution that
can be tested by further experiments.

\begin{appendix}
\section{The impact of time sampling on the performance of HM-NEMs}\label{app-timeSamp}
\begin{figure}[t]

\includegraphics{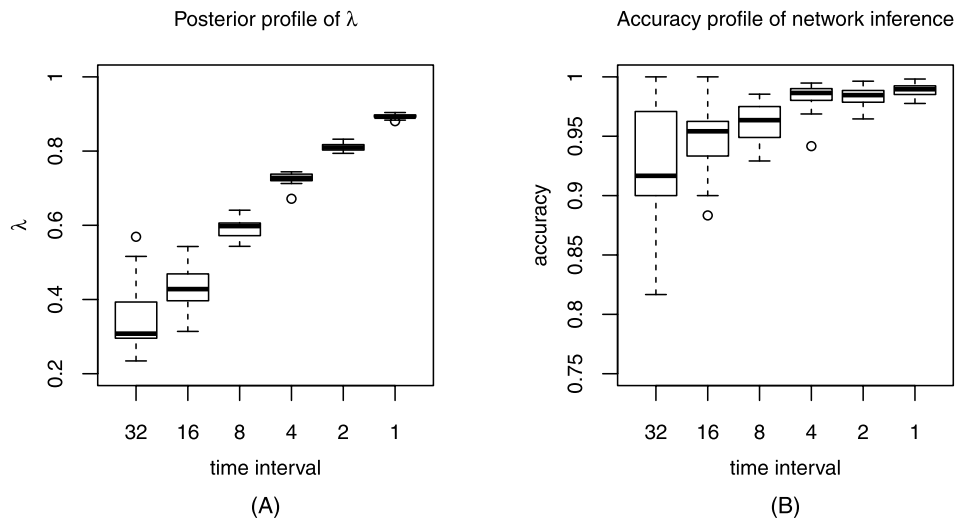}

\caption{\textup{(A)} Posterior profile of $\lambda$, and \textup{(B)}
accuracy profile of network inference
as a function of the interval of time sampling.}\label{simtimesamp}
\end{figure}

Without loss of generality, we set $\lambda=0.9$ and generated a
random evolving
network of $n=6$ pathway components over $T=128$ time points. Setting
the same parameters
($n_r=4$ reporters per pathway component, $n_p=3$ replicates, $\alpha
=0.1$ and $\beta=0.1$)
as the simulation study in Section~\ref{secconv}, we generated an
artificial perturbation
data set. Next, we sampled the complete data set with the time interval\vadjust{\goodbreak} varying
from 1 to 32. HM-NEMs are then applied to estimate the posterior of
$\lambda$ and evolving network for each setting of time sampling. As expected,
the posterior mean of $\lambda$ increases and approaches the real
$\lambda$ as
time interval decreases [Figure~\ref{simtimesamp}(A)], indicating
that a smaller time interval
tends to give a ``smoother'' estimate of the network. The performance of network
inference, assessed by accuracy, is also improved by more time samples
[Figure~\ref{simtimesamp}(B)]. However, even for only 4 time points
(time interval${}=32$),
HM-NEMs achieved a good performance (median accuracy${}=0.92$).

%

%
\begin{figure}[t]

\includegraphics{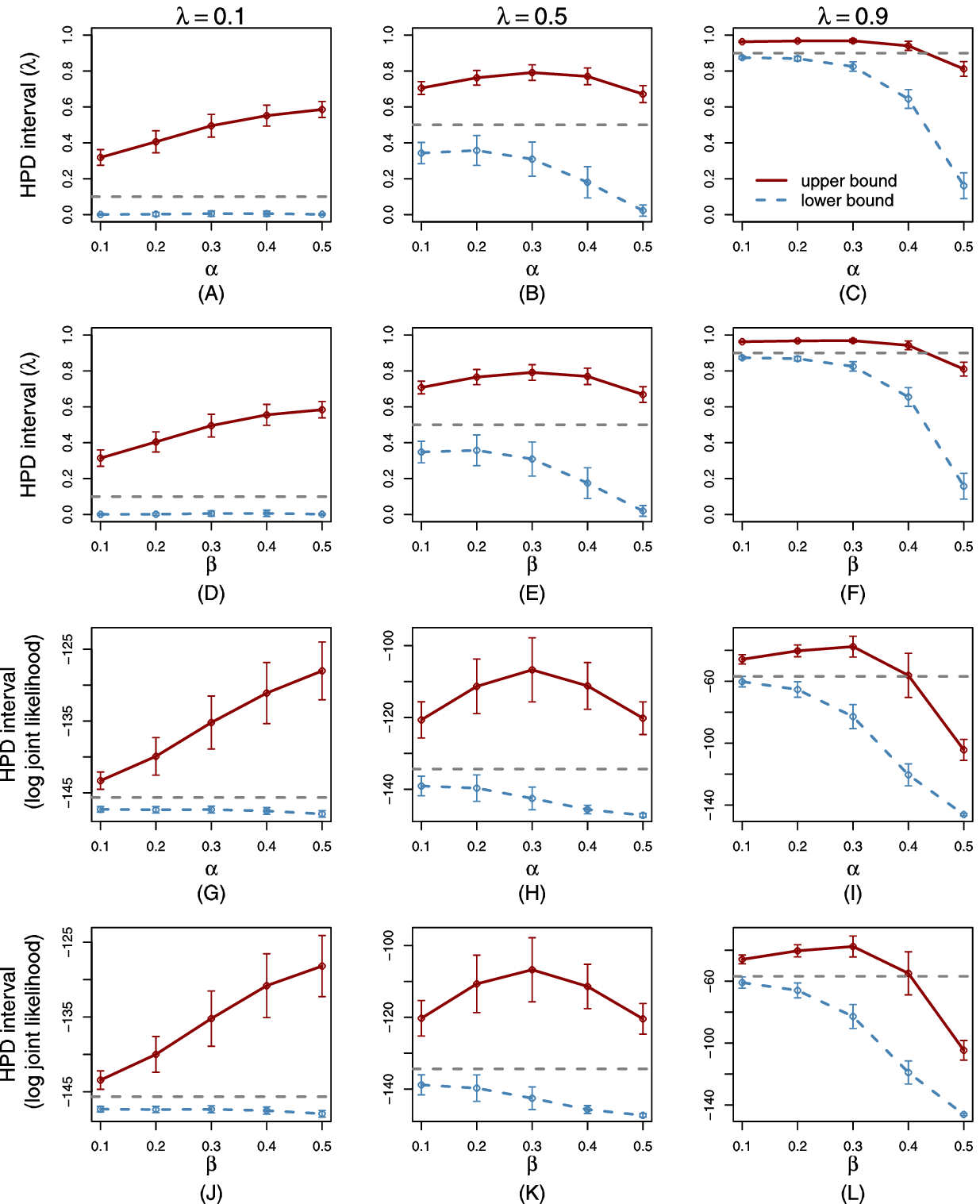}

\caption{HPD intervals of $\lambda$ [\textup{(A)} to \textup{(F)}] and log joint likelihood
[\textup{(G)} to \textup{(L)}] as a function of
$\alpha$ (when $\beta=0.3$) and $\beta$ (when $\alpha=0.3$).
Solid red and dashed blue lines represent upper and lower confidence
bounds, respectively.
Error bars indicate one standard deviation of the mean.
The true $\lambda$ and log joint likelihood are denoted by dashed gray lines.}\label{simintervals}
\end{figure}

\section{HPD intervals as a function of $\alpha$ and $\beta$}\label{app}

For each combination of parameter $\alpha$, $\beta$ and $\lambda$, we
computed highest posterior density (HPD) intervals with 95\% nominal
coverage probability for estimated $\lambda$ and log joint likelihood.
As shown in Figure~\ref{simintervals}, HPD intervals for estimated
$\lambda$ and log joint likelihood both cover the true $\lambda$ and
likelihood even when $\alpha$ and $\beta$ are as high as 30\%,
demonstrating the robustness of our algorithm to noise in observed perturbation
data.
\end{appendix}


\section*{Acknowledgements}
We thank Professor Steve Altschuler and Dr. Chin-Jen Ku at the
University of
Texas Southwestern Medical Center for kindly sharing the neutrophil
perturbation data. We also thank Dr. Achim Tresch at the Max Planck
Institute in Cologne and Dr. Roland F. Schwarz at the European
Bioinformatics Institute for helpful discussions and commenting on
drafts of the paper.


%

\printaddresses

\end{document}